%% file: neurips_2025.tex
\PassOptionsToPackage{numbers, compress}{natbib}
\documentclass{article}
\usepackage[preprint]{neurips_2025}

\usepackage[utf8]{inputenc} % allow utf-8 input
\usepackage[T1]{fontenc}    % use 8-bit T1 fonts     % hyperlinks
\usepackage{url}            % simple URL typesetting
\usepackage{booktabs}       % professional-quality tables
\usepackage{amsfonts}       % blackboard math symbols
\usepackage{nicefrac}       % compact symbols for 1/2
\usepackage{microtype}  
\usepackage{wrapfig}
\usepackage{multicol}
\usepackage{multirow}
\usepackage{booktabs}
\usepackage{amsmath,amsfonts}
\usepackage{algorithmic}
\usepackage{algorithm}
\usepackage{graphicx}
\usepackage{enumitem}
\usepackage{xspace}
\usepackage{caption}
\usepackage{pifont}
\usepackage{subcaption}
\usepackage{natbib}
\usepackage{colortbl}
\usepackage{xcolor}         % colors
\usepackage[colorlinks,
            linkcolor=red,
            anchorcolor=blue,
            citecolor=cyan
            ]{hyperref}
\newcommand{\ie}{\emph{i.e., }}
\newcommand{\eg}{\emph{e.g., }}

\newcommand{\wrt}{\emph{w.r.t. }}
\newcommand{\cf}{\emph{cf. }}

\title{On Negative-aware Preference Optimization for Recommendation}

\author{
Chenlu Ding\thanks{These authors contributed equally to this work.} \\
USTC \\
\texttt{dingchenlu200103@gmail.com}
\And
Daoxuan Liu\footnotemark[1] \\
USTC \\
\texttt{daoxuanliu6274@gmail.com}
\And
Jiancan Wu \\
USTC \\
\texttt{wujcan@gmail.com}
\AND
Xingyu Hu \\
USTC \\
\texttt{huxy@mail.ustc.edu.cn}
\And
Junkang Wu \\
USTC \\
\texttt{jkwu0909@gmail.com}
\And
Haitao Wang \\
Meituan \\
\texttt{wanghaitao13@meituan.com}
\AND
Yongkang Wang \\
Meituan \\
\texttt{wangyongkang03@meituan.com}
\And
Xingxing Wang \\
Meituan \\
\texttt{wangxingxing04@meituan.com}
\And
Xiang Wang \\
USTC \\
\texttt{xiangwang1223@gmail.com}
}

\begin{document}
\maketitle
\input{chapters/0_abs}
\input{chapters/1_intro}
\input{chapters/2_pre}

\input{chapters/3_method}

\input{chapters/4_exp}

\input{chapters/6_conclusion}

\bibliographystyle{unsrt}
\bibliography{refnip}

\input{chapters/7_appendix}

\end{document}

%% file: chapters/0_abs.tex
\begin{abstract}
Recommendation systems leverage user interaction data to suggest relevant items while filtering out irrelevant (negative) ones. The rise of large language models (LLMs) has garnered increasing attention for their potential in recommendation tasks. However, existing methods for optimizing LLM-based recommenders face challenges in effectively utilizing negative samples. Simply integrating large numbers of negative samples can improve ranking accuracy and mitigate popularity bias but often leads to increased computational overhead and memory costs. Additionally, current approaches fail to account for the varying informativeness of negative samples, leading to suboptimal optimization performance. To address these issues, we propose NAPO (\textbf{N}egative-\textbf{A}ware \textbf{P}reference \textbf{O}ptimization), an enhanced framework for preference optimization in LLM-based recommendation. NAPO introduces two key innovations: (1) in-batch negative sharing, which expands the pool of negative samples without additional memory overhead, and (2) dynamic reward margin adjustment, which adapts model updates based on the confidence of negative samples. Extensive experiments on three public datasets demonstrate that NAPO outperforms existing methods in both recommendation accuracy and popularity bias reduction.
\end{abstract}

%% file: chapters/1_intro.tex
\section{Introduction}
Recommendation systems leverage user interaction data to suggest items that users are likely to interact with, while avoiding those out of interest (\ie negative items). With the emergence of large language models (LLMs) \cite{DBLP:journals/corr/abs-2407-21783,DBLP:journals/corr/abs-2303-08774}, post-training LLMs to serve as recommenders has gained significant attention \cite{DBLP:journals/corr/abs-2305-07622,DBLP:conf/recsys/XiLLCZZCT0024,sheng2024language,DBLP:conf/cvpr/LiuLLL24,DBLP:conf/iclr/WeiBZGYLDDL22}. 
The supervision signals for LLM post-training are predominantly derived from the target response (\ie the positive item), with negative items sampled from non-observed interaction space only appearing in input prompt.
Such an approach may obscure nuanced distinctions between positive and negative preferences, making it challenging for the LLM to effectively differentiate between them \cite{wang2023generative,hou2024large}.

Direct Preference Optimization (DPO) \cite{DBLP:conf/nips/RafailovSMMEF23} emerges as an effective method for aligning LLMs with human values.
It shares similar foundations with the classic Bayesian Personalized Ranking (BPR) \cite{DBLP:conf/uai/RendleFGS09}, maximizing the reward difference between positive and negative responses, with a reference model serving as an optimization anchor.
Follow-on work, SimPO \cite{DBLP:journals/corr/abs-2405-14734}, simplifies the objective formulation by eliminating the reference model requirement while introducing a reward margin to control the separation between positive and negative responses.
Recently, researchers have explored various DPO variants tailored for LLM-based recommenders \cite{DBLP:journals/corr/abs-2406-09215,DBLP:journals/corr/abs-2410-12519,gao2024sprec,DBLP:journals/corr/abs-2405-16127}, enabling models to distinguish between preferred and irrelevant items in the vast item space.
These efforts primarily focus on better utilization of negative samples, ranging from selecting semantically similar negatives for establishing clear decision boundaries \cite{DBLP:journals/corr/abs-2410-12519}, to incorporating larger negative pools for more comprehensive supervisory signals \cite{DBLP:journals/corr/abs-2405-16127,DBLP:journals/corr/abs-2406-09215}.

\begin{figure}[t]
  \centering
  \begin{minipage}[b]{0.69\textwidth}
    \centering    \includegraphics[width=\textwidth]{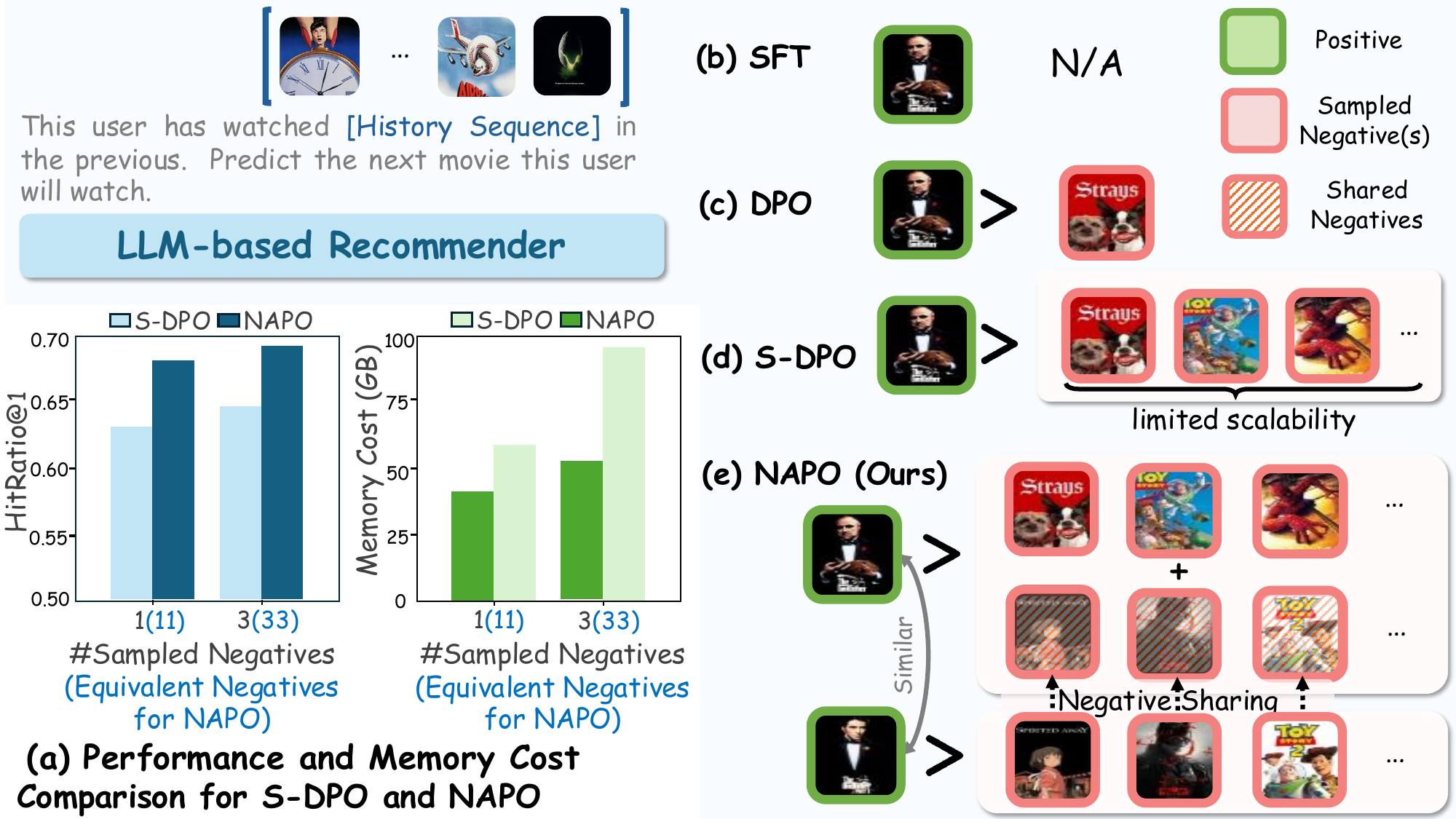}
    \vspace{-15pt}
  \end{minipage}
  \begin{minipage}[b]{0.3\textwidth}
    \centering    \includegraphics[width=\textwidth]{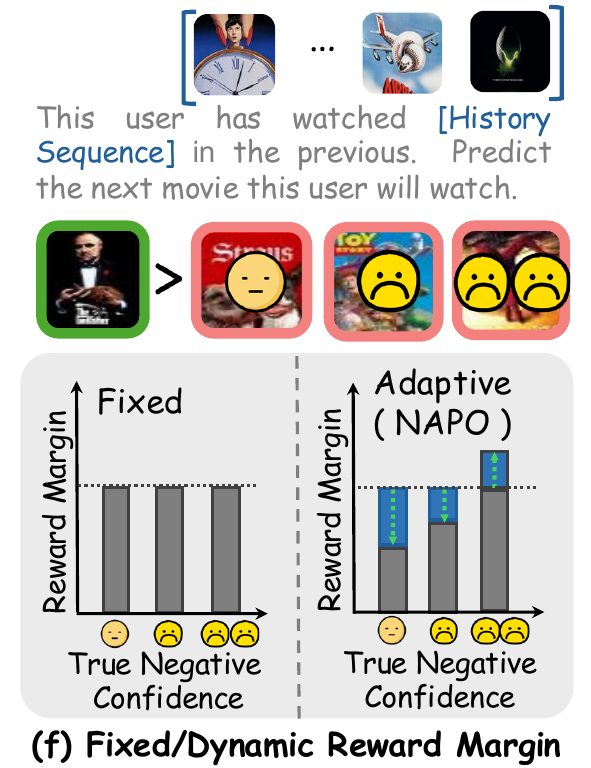}
    \vspace{-15pt}
\end{minipage}
  \caption{Comparison of Supervised Fine-tuning (SFT) and DPO-based methods in negative utilization and reward margin control. (a) Performance and memory cost of S-DPO and NAPO. "Equivalent Negatives FOR NAPO" refers to the effective number of negative samples utilized in NAPO, expanded through its in-batch negative sharing strategy. (b)-(e) Different negative utilization strategies: SFT, DPO, SDPO, and NAPO. (f) Fixed vs. Adaptive reward margin.}
  \label{intro}
  \vspace{-15pt}
\end{figure}

Despite effectiveness, we argue that two significant challenges persist in
effectively exploiting negative signals for optimizing DPO-based recommenders:

\begin{itemize}[leftmargin=*]
 \vspace{-7pt}
    \item \textbf{Efficient Integration of More Negative Signals}:  In DPO-based recommendation, effectively leveraging the vast unexplored negative sample space is crucial for improving ranking accuracy (Figure~\ref{intro}(a)) and debiasing \cite{wu2020sgl,wu2024ssm,DBLP:journals/corr/abs-2406-09215,DBLP:conf/cikm/MaoZWDDXH21}. Ensuring a larger set of negatives provides richer supervision, enabling the model to better differentiate between relevant and irrelevant items.
    % In DPO-based recommendation, the supervision signals from positive samples are inherently limited, while the negative sample space offers vast unexplored potential. Ensuring a larger set of negatives is crucial for improving ranking accuracy (Figure~\ref{intro}(a)) and debiasing. 
    However, expanding the negative sample pool comes at a high computational cost, as illustrated in Figure~\ref{intro}(a). Unlike traditional negative sample sharing, where negative embeddings are precomputed and reused, LLM-based recommenders must decode each negative sample based on the user context within the prompt. Each additional negative requires separate decoding computations, significantly increasing training time and memory usage. Moreover, naive integration of numerous negative samples may introduce false negative examples, diluting the optimization signal and even degrading performance \cite{ma2024negative,yang2020mixed}. Effective integration requires a scalable approach that balances the quantity and quality of negative samples while maintaining efficiency.
     \vspace{-2pt}
    \item \textbf{Context-Aware Negative Sample Optimization}:
    Traditional DPO methods, including SimPO, rely on a fixed reward margin to separate positive and negative samples, implicitly treating all negatives as equally important. However, this static design fails to account for the varying informativeness of negative samples \cite{yang2020mixed,zhou2021contrastive}. Specifically, semantically similar negative items, which are close to positives in representation space, require more cautious updates. These negatives pose a higher risk of conflicting with positive signals, potentially leading to optimization instability \cite{ma2024negative,yang2020mixed}. Ignoring this distinction can lead to over-penalizing valuable negatives or under-emphasizing truly irrelevant ones, distorting the model's learning process \cite{bruch2019revisiting,yang2020mixed}. 
     \vspace{-7pt}
\end{itemize}

To address these challenges, we propose \textbf{NAPO} (\textbf{N}egative-\textbf{A}ware \textbf{P}reference \textbf{O}ptimization for Recommendation), to enhance negative sample utilization.
We adopt SimPO's formulation \cite{DBLP:journals/corr/abs-2405-14734}, which simplifies the preference optimization objective by directly using the language model's log probabilities while eliminating the need for the reference model.  Building on this, we propose two innovations for DPO-based recommendation: an \textit{in-batch negative sharing strategy}, and a \textit{dynamic impact assessing mechanism}.

\begin{itemize}[leftmargin=*]
 \vspace{-7pt}
    \item For \textit{negative sharing}, we leverage the model’s existing computation of log probabilities during standard forward passes. Instead of independently decoding each negative sample, we establish a batch-wise sharing mechanism where the computed log probabilities of negatives are dynamically shared among other instances within the same batch, effectively forming a shared negative pool without incurring additional computational overhead\footnote{In distributed training, negative probability sharing occurs at the batch level, exchanging similarity-guided signals across nodes.}. However, the validity of shared negatives hinges on the similarity of user interests --- a negative meaningful for one sequence may be irrelevant to another. We employ a similarity-guided mechanism that filters negative sharing using sequence embeddings from a lightweight model (\eg SASRec \cite{DBLP:conf/icdm/KangM18}). Negatives are shared only between sequence pairs with top similarity scores, maintaining contextual relevance.
    \item While for \textit{dynamic impact assessing mechanism}, we extend beyond SimPO's fixed reward margin and propose a dynamic reward margin mechanism, which adaptively adjusts the reward gap between positives and their negative counterpart, as shown in Figure~\ref{intro}(f). Specifically, it scales the impact of different negatives on model updates based on the confidence level predicted by the auxiliary lightweight sequential recommender. For high-confidence negative samples strongly supported by signals from traditional recommenders, we amplify their impact by applying a larger margin. Conversely, for low-confidence negative samples, where the certainty of an item being negative is reduced, we apply a smaller margin to encourage more cautious updates. This confidence-driven adjustment ensures the model emphasizes reliable negatives, enhancing preference optimization and overall performance.
    \vspace{-10pt}
\end{itemize}

Extensive experiments on $3$ public recommendation datasets (\ie Goodreads\footnote{\url{https://www.goodreads.com.}}, LastFM \cite{DBLP:conf/recsys/2011hetrec}, Steam \cite{DBLP:conf/icdm/KangM18}) demonstrate that NAPO effectively improves recommendation performance by $13\%$ while significantly reducing popularity bias. Notably, NAPO leverages a substantially larger set of negative samples compared to existing methods, without increasing memory or computational overhead.
% Notably, under the same memory constraints and comparable computational overhead, NAPO achieves an equivalent negative sampling capacity of $209$ samples - a $13.9\times$  improvement over existing method \cite{DBLP:journals/corr/abs-2406-09215} which is strictly capped at $15$ samples.

%% file: chapters/2_pre.tex
\section{Preliminary}
% This section introduces DPO and SimPO as foundational methods for aligning LLMs with human preferences, followed by a discussion on fine-tuning LLMs for recommendation using supervised fine-tuning (SFT) and preference optimization.

\subsection{Direct Preference Optimization}
\subsubsection{\textbf{DPO}} Direct Preference Optimization (DPO) \cite{DBLP:conf/nips/RafailovSMMEF23} is a method for aligning LLMs with human values, building on Reinforcement Learning with Human Feedback (RLHF) \cite{DBLP:conf/nips/Ouyang0JAWMZASR22,DBLP:conf/nips/ChristianoLBMLA17}. It optimizes user preferences using pairwise data, consisting of a preferred response $y^{+}$ and a less-preferred response $y^{-}$ to a prompt $x$, forming a dataset $\mathcal{D} = {(x, y^{+}, y^{-})}$. The preference between $y^{+}$ and $y^{-}$ is modeled using the Bradley-Terry (BT) model \cite{bradley1952rank}: $p_{\text{DPO}}(y^{+} \succ y^{-} | x) = \sigma(r(x,y^{+}) - r(x,y^{-})),$
where $r(x,y)$ represents the reward function quantifying the alignment of $y$ with human preferences in the context of a given prompt $x$, and $\sigma$ is the sigmoid function.

The policy model $\pi_{\theta}(y|x)$, parameterized by $\theta$, predicts the probability distribution over possible responses $y$ given a prompt $x$.
The objective of DPO is to fine-tune policy model to align with user preferences by maximizing the likelihood of preferred responses while penalizing less-preferred ones:
\begin{equation}
\label{maxtarget}
\max _\theta \mathbb{E}_{\left(x, y^{+}, y^{-}\right) \sim \mathcal{D}} \log p_{\text{DPO}}\left(y^{+} \succ y^{-} \mid x\right).
\end{equation}
Rather than explicitly learning a reward model, DPO \cite{DBLP:conf/nips/RafailovSMMEF23} directly derives the reward function as:
\begin{equation}
\begin{aligned}
    r(x,y) = \hat{H}_\theta(y \mid x)+\beta \log Z(x),
\end{aligned}
\end{equation}
where $\hat{H}_\theta(y \mid x) = \beta \log \left( \frac{\pi_{\theta}(y \mid x)}{\pi_{\text{ref}}(y \mid x)} \right)$. Here $\pi_{\text{ref}}$ is the reference model (\ie a frozen pre-trained LLM) that provides a baseline policy to ensure the alignment improvements over an established standard. $Z(x) = \sum_y \pi_{\text{ref}}(y \mid x) \exp \left(\frac{1}{\beta} r(x, y)\right)$ is the partition function.

Incorporating this reward into the objective, DPO's optimization becomes:
\begin{equation}
\begin{aligned}
\mathcal{L}_{\text{DPO}} =  -\mathbb{E}_{( x, y^{+}, y^{-})\sim \mathcal{D}} \Big[ \log \sigma \left( \hat{H}_\theta(y^+ \mid x)  - \hat{H}_\theta(y^- \mid x) \right) \Big],
\end{aligned}
\end{equation}
where $\beta$ balances the policy model and the reference model, controlling how closely the optimized model aligns with the reference.

\subsubsection{\textbf{SimPO}} Unlike original DPO, SimPO \cite{DBLP:journals/corr/abs-2405-14734} discards the reference model from the reward function, relying solely on the policy model's average log-likelihood: 
    \begin{align}
    r(x,y) = \frac{1}{|y|} H_\theta(y \mid x),
\end{align}
where $H_\theta(y \mid x) = \beta \log \pi_{\theta}(y \mid x)$. This eliminates the computational and memory overhead associated with maintaining a reference model, making the approach more efficient. Additionally, SimPO introduces a fixed reward margin $\gamma_0$ to ensure a sufficient reward difference between the chosen and rejected responses. The likelihood ${p}_{\text{SimPO}}(y^{+} \succ y^{-} | {x})$ can be written as: 
\begin{align}
\label{reward_gamma}
{p}_{\text{SimPO}}(y^{+} \succ y^{-} | {x}) & = \sigma(r(y^{+} , {x}) - r(y^{-}, {x}) - \gamma_0) \\ & =\sigma\Big(\frac{1}{\left|y^{+}\right|} H_\theta(y^+ \mid x) - \frac{1}{\left|y^{-}\right|} H_\theta(y^- \mid x) - \gamma_0\Big).\nonumber
\end{align}
The optimization objective for SimPO is:
\begin{equation}
\mathcal{L}_{\text{SimPO}} =  -  \mathbb{E}_{( x, y^{+}, y^{-})\sim \mathcal{D}}  \Bigg[ \log \sigma \Bigg( \frac{1}{\left|y^{+}\right|} H_\theta(y^+ \mid x)   - \frac{1}{\left|y^{-}\right|} H_\theta(y^- \mid x) - \gamma_0 \Bigg) \Bigg].
\end{equation}

\subsection{Fine-tuning of LLM-based Recommenders}
Given a sequence of historical items $s_u = \{i_1, i_2, ..., i_n\}$ that user $u$ interacted with, sequential recommendation predicts the next likely item. Training LLMs as sequential recommenders typically involves two major components: supervised fine-tuning (SFT) and preference optimization.

\subsubsection{\textbf{Supervised Fine-tuning}}
Leading LLM-based recommenders \cite{DBLP:conf/recsys/BaoZZWF023,DBLP:journals/corr/abs-2408-10159,DBLP:journals/corr/abs-2312-02445} predominantly adopt SFT on interaction data to enhance their recommendation ability.
It first converts the interaction data into a text-format dataset $\mathcal{D}_\text{SFT} = \{(x_u, y_u^{+})\}$, with the textual prompt $x_u$ expressed as:
\begin{equation}
x_u={\text{Prompt}}\left(X, s_u\right),
\end{equation}
where $X$ represents the textual instruction for the recommendation task and the candidate item list. $y^{+}$ is the textual response describing the target item within the candidates that the user is expected to interact with subsequently.
Then, it uses autoregressive language modeling to maximize the probability $\pi_{\phi}(y_u^{+} \mid x_u)$, where $\pi_{\phi}$ represents the probability predicted by the recommender with trainable parameters $\phi$ (\eg LoRA \cite{DBLP:journals/corr/abs-2106-09685} parameters). The optimization objective is:
\begin{equation}
\max _\phi \mathbb{E}_{(x_u, y_u^{+}) \sim \mathcal{D}_{\mathrm{SFT}}} \pi_\phi(y_u^{+}\mid x_u).
\end{equation}
However, this approach mainly derives supervision from the target response $y_u^{+}$, while negative samples are simply embedded in the input prompt $x_u$. This can obscure nuanced distinctions between positive and negative preferences, limiting the model's ability to fully exploit interaction data.

\subsubsection{\textbf{Preference Optimization for Recommendation}}
To address the limitations of SFT, an additional preference optimization stage is introduced. This stage explicitly integrates negative samples as supervision signals, enabling the model to distinguish between positive and negative items in the interaction space. S-DPO \cite{DBLP:journals/corr/abs-2406-09215} is a pioneering approach using the Plackett-Luce (PL) model \cite{plackett1975analysis,debreu1960individual} to model preference distributions with multiple negatives. Given a prompt $x_u$ representing user $u$'s interaction history and the positive item $y_u^{+}$, it samples a set of negative items $\hat{\mathcal{E}_u}$ from the unobserved space, constructing $y^-$. The multi-negative preference distribution $\hat{p}$ is then defined as:
\begin{equation}
    \hat{p}(y_u^{+} \succ_u \hat{\mathcal{E}_u} \mid x_u) = \frac{\exp(\hat{H}_\theta(y^+_u \mid x_u))}{\sum_{y_u^{-} \in \hat{\mathcal{E}_u}} \exp(\hat{H}_\theta(y^-_u\mid x_u))},
\label{sdpo_distribution}
\end{equation}
% where $n_{neg}$ is the number of negative samples. 
where $y_u^{+} \succ_u \hat{\mathcal{E}_u}$ means $y^+_u \succ y^-_u$ for $\forall y_u^- \in \hat{\mathcal{E}_u}$. The loss of S-DPO can therefore be expressed as:
\begin{align}
\mathcal{L}_{\text{S-DPO}}
= -\mathbb{E}&_{( x_u, y_u^{+}, \hat{\mathcal{E}_u}) \sim \mathcal{D}} \Bigg[ 
    \log \sigma \left( 
        \log \hat{p}(y_u^{+} \succ_u \hat{\mathcal{E}_u} \mid x_u)\right)
\Bigg].
\label{eq:sdpo-loss}
\end{align}
Here, $\pi_{\text{ref}}$ is the reference model, typically the pre-trained SFT model. By applying preference optimization after SFT, S-DPO can better capture nuanced user preferences, enhancing recommendation performance.

%% file: chapters/3_method.tex
\section{Methodology}
We introduce \textbf{NAPO}, a \textbf{n}egative-\textbf{a}ware \textbf{p}reference \textbf{o}ptimization framework for enhancing negative signal utilization in LLM-based recommenders. Section \ref{ssec:overview} outlines NAPO's optimization objective, focusing on two key components: in-batch negative sharing and dynamic impact assessment. Section \ref{ssec:in-batch} describes the in-batch negative sharing strategy, which expands negative coverage without extra computation. Section \ref{ssec:dynamic_gamma} details the dynamic reward margin mechanism, adaptively adjusting negative sample impact based on confidence.

\subsection{Overview of NAPO}
\label{ssec:overview}
In LLM-based recommender models, the effective utilization of unobserved interactions is crucial for distinguishing between positive and negative items \cite{DBLP:journals/corr/abs-2406-09215,DBLP:journals/corr/abs-2410-12519}.
While integrating more negative signals can help mitigate popularity bias \cite{wu2020sgl,wu2024ssm} and enhance recommendation accuracy \cite{DBLP:journals/corr/abs-2406-09215,DBLP:conf/cikm/MaoZWDDXH21}, leveraging the abundant negative samples presents significant challenges on efficiency and computational costs.
To bridge this research gap, we propose NAPO, a framework that introduces two key designs: an in-batch negative sharing strategy and a dynamic impact assessment mechanism.
Formally, the optimization objective of NAPO can be expressed as:
\begin{equation}
  \mathcal{L}_{\text{NAPO}}
= -\mathbb{E}_{(x_u, y_u^{+}, \mathcal{E}_u) \sim \mathcal{D}} \Bigg[
    \log \sigma \Big( \log {p}(y_u^{+} \succ_u {\mathcal{E}_u} \mid x_u) - \gamma
    \Big)
\Bigg],
\label{eq:napo_loss}
\end{equation}
% \begin{equation}
%   \mathcal{L}_{\text{NAPO}}
% = -\mathbb{E}_{(x_u, y_u^{+}, \mathcal{E}_u) \sim \mathcal{D}} [
%     \log \sigma ( \log {p}(y_u^{+} \succ_u {\mathcal{E}_u} \mid x_u) - \gamma
%     )],
% \label{eq:napo_loss}
% \end{equation}
where ${p}(y_u^{+} \succ_u {\mathcal{E}_u} \mid x_u)$ represents the probability of preferring the positive item over all negative ones in the hybrid negative item set $\mathcal{E}_u$, expressed as:
\begin{equation}
{p}(y_u^{+} \succ_u {\mathcal{E}_u} \mid x_u) = \frac{\exp({H}_\theta(y^+_u \mid x_u))}{\sum_{y_u^{-} \in {\mathcal{E}_u}} \exp({H}_\theta(y^-_u\mid x_u))},
\end{equation}
% where $\mathcal{E}_u$ is the hybrid negative sample set, which includes both the original sampled negative set $\hat{\mathcal{E}_u}$ and the in-batch shared negative samples.
Our proposed NAPO differs from current methods in the following three key aspects:
\begin{itemize}[leftmargin=*]
\vspace{-7pt}
    \item \textbf{Simplified Probability Choice}: Inspired by SimPO \cite{DBLP:journals/corr/abs-2405-14734}, NAPO utilizes the scaled log-probability $H_{\theta} (y|x)$ in place of $\hat{H}_{\theta} (y|x)$ used in S-DPO.
    This modification eliminates the need for a reference model, reducing computational overhead \wrt the reference model and achieving better alignment with generative tasks \cite{DBLP:journals/corr/abs-2405-14734,DBLP:journals/corr/abs-2410-10148}.
    \item \textbf{Hybrid Negative Set}: NAPO enlarges the negative set to a hybrid collection, including both random negatives $\hat{\mathcal{E}}$ sampled for each sequence and in-batch shared negative samples $\mathcal{S}$,
    % enabling the better use of the abundant negative signals.
    enhancing the negative sample coverage without incurring additional computational costs.
    
    \item \textbf{Dynamic Reward Margin}: NAPO extends the fixed margin coefficient adopted in SimPO, introducing a dynamic margin, $\gamma$, which adapts to user interests, providing a more nuanced assessment of the negative samples' impact.
\vspace{-7pt}
\end{itemize}

For a clearer comparison, we summarize the DPO-based objectives for recommendation in Table \ref{tab:loss_compare}.

\begin{table}[t]
\vspace{-7pt}
\caption{A summary of different preference alignment methods.}
\small
\label{tab:loss_compare}
\resizebox{\textwidth}{!}{\begin{tabular}{ccccl}
\toprule
\textbf{Method}  & \textbf{\begin{tabular}[c]{@{}c@{}}w/o Reference  Policy\end{tabular}} & \textbf{\begin{tabular}[c]{@{}c@{}}\#Negative  Samples\end{tabular}} & \textbf{\begin{tabular}[c] {@{}c@{}}Reward Margin\end{tabular}} & \multicolumn{1}{c}{\textbf{Objective}} \\ \midrule
DPO \cite{DBLP:conf/nips/RafailovSMMEF23}       &      \textcolor{red}{\ding{55}}    & 1                                                                    & -                                                                &      $-  \log \sigma\left(\log p_{\text{DPO}}(y^{+}_u \succ y^{-}_u | x_u)\right)$           \\
SimPO \cite{DBLP:journals/corr/abs-2405-14734}    &    \textcolor{green}{\checkmark}                                                                         & 1                                                                    & fixed                                                        &   $- \left[ \log \sigma\left(\log p_{\text{SimPO}}(y^{+}_u \succ y^{-}_u | x_u)\right) - \gamma_0 \right]$        \\
S-DPO \cite{DBLP:journals/corr/abs-2406-09215}        &     \textcolor{red}{\ding{55}}      & mutilple                                                             & -                             &    $-  \log \sigma\left(\log \hat{p}(y^{+}_u \succ_u \hat{\mathcal{E}_u} | x_u)\right)$                                            \\
RosePO \cite{DBLP:journals/corr/abs-2410-12519}      &     \textcolor{red}{\ding{55}}    & 1                                                                    & -                                                  &    $-  \ (1-\epsilon_\phi) \log p_{\text{DPO}}(y^{+}_u \succ y^{-}_u | x_u) - \epsilon_\phi \log p_{\text{DPO}}(y^{-}_u \succ y^{+}_u | x_u)  $                         \\

$\alpha$-DPO \cite{DBLP:journals/corr/abs-2410-10148}        &    \textcolor{green}{\checkmark}      & 1                                                             & {dynamic}                             &    $-  \left[ \log \sigma \left( u(x_u, y^+_u, y^-_u) \right) - sg\left( \gamma_0 + \alpha M^*(x_u, y^+_u, y^-_u) \right) \right]
$                                            \\

\textbf{NAPO}     &       \textcolor{green}{\checkmark}                                                                        & \textbf{hybrid}                                                               & \textbf{dynamic}                                    &    $-  \left[  \log \sigma\left(\log {p}(y^{+}_u \succ_u \mathcal{E}_u | x_u)\right) - \gamma \right]$                             \\ \bottomrule
\vspace{-15pt}
\end{tabular}}
\end{table}

\subsection{In-batch Negative Sharing}

\begin{figure}[t]
  \centering
  \begin{minipage}[b]{0.99\textwidth}
    \centering    \includegraphics[width=\textwidth]{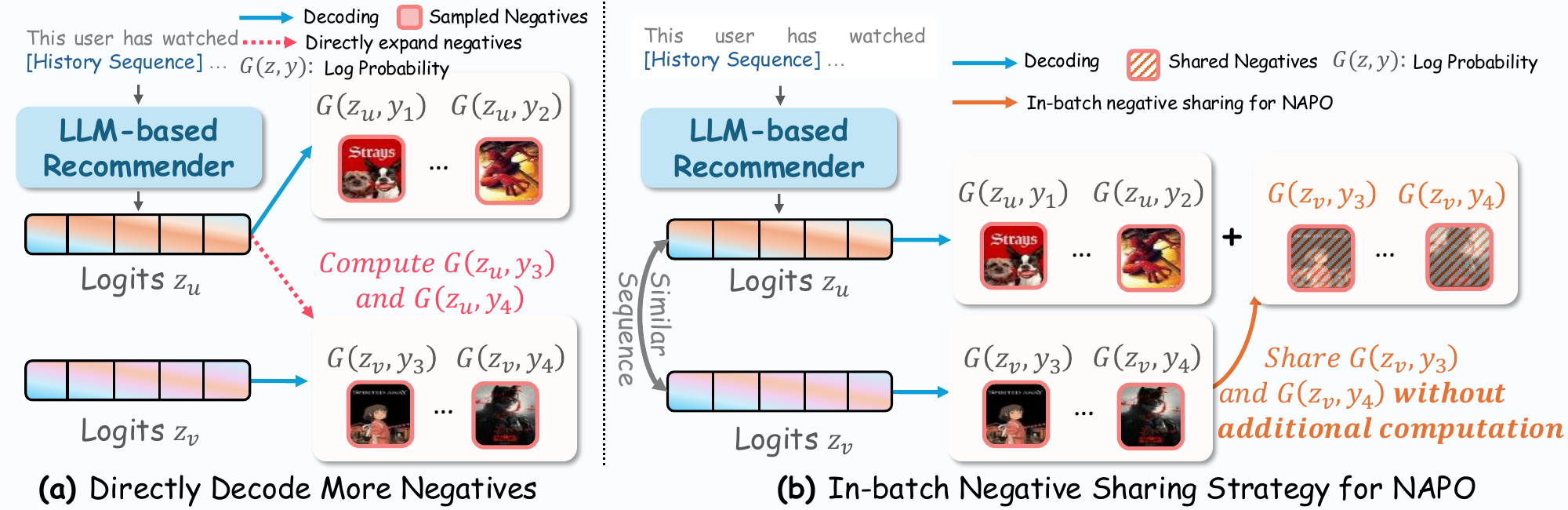}
    \vspace{-15pt}
  \end{minipage}
  \caption{Comparison of different methods expanding negative coverage in DPO-based recommenders. (a) Each sequence computes $G(z_u, y_v^-)$ for all negative samples (b) $G(z_u, y_v^-)$ are shared among similar sequences without extra computation.
}
  \label{method}
  \vspace{-20pt}
\end{figure}

\label{ssec:in-batch}
% Let $\mathcal{B} = \{ (x_1,y_1^+, \hat{\mathcal{E}_1}), (x_2,y_2^+, \hat{\mathcal{E}_2}), \cdots, (x_n,y_n^+, \hat{\mathcal{E}_n}) \}$ denote a batch of training instances.
Let $\mathcal{B} = \{ (x_u,y_u^+, \hat{\mathcal{E}_u})\}$ denote a batch of training instances.
Each triplet $(x_u,y_u^+, \hat{\mathcal{E}_u})$ consists of three elements:
$x_u$ represents a prompt comprising the recommendation task instruction and the user's historical interaction sequence $s_u$,
$y_u^{+}$ is the target item to be predicted,
and $\hat{\mathcal{E}_u}$ denotes the set of negative items randomly sampled from the pool of unobserved interactions.

Scrutinizing the computation process of the scale log probability $H_{\theta} (y | x)$ in LLM-based recommenders, we recognize that it can be naturally decomposed into two functions:
\begin{equation}
    H_{\theta}(y_u | x_u) = G( F (x_u), y_u), \quad y_u \in \{ y_u^{+}\} \cup \hat{\mathcal{E}_u},
\end{equation}
where $F: X \rightarrow \mathbb{R}^{d}$ first maps the input prompt $x_u \in \mathcal{X}$ to intermediate logits $z_u \in \mathbb{R}^{d}$ through multiple Transformer layers \cite{DBLP:conf/nips/VaswaniSPUJGKP17}, and $G$ then processes these logits autoregressively to generate the final scaled logarithmic probability.
This decomposition suggests dual computational bottlenecks --- function $F$ requires multiple passes through the large-scale Transformer architecture, while function $G$ needs autoregressive decoding for each $(x_u, y_u)$ pair. Unlike traditional methods that share negative item embeddings directly \cite{wu2020sgl}, DPO-based optimization relies on decoding each negative sample (function $G$) based on the logits. Thus the decomposition of prompt-dependent logits generation (function $F$) and negative item processing (function $G$) necessitates a different strategy.
% This two-stage computational process fundamentally differentiates negative sharing in LLM-based recommenders from traditional approaches. While traditional approaches can directly share item embeddings \cite{wu2020sgl}, the decomposition of prompt-dependent logits generation (function $F$) and negative item processing (function $G$) necessitates a different strategy.

An intuitive approach to expanding negative samples in LLM-based recommenders is to directly compute $G(z_u, y_v^{-})$ for each user sequence with a wide range of negative items, as shown in Figure~\ref{method}(a). Specifically, for each sequence $u$, the model decodes logits $z_u$ for the positive item and a set of sampled negatives. To further increase negative coverage, $G(z_u, y_v^{-})$ are computed for all negative items associated with any other sequence in the batch. However, this approach suffers from high computational cost and memory overhead due to repeated computation and storage of $G(z_u, y_v^{-})$ of $n \lvert \hat{\mathcal{E}} \rvert$ negatives within a  batch. The complexity scales as
$O(n^2 \lvert \hat{\mathcal{E}} \rvert)$, where $n$ is the batch size and $\lvert \hat{\mathcal{E}} \rvert$ is the number of sampled negatives per sequence. 
% Similarly, while S-DPO demonstrates benefits from increased negative samples, its requirement to recompute both functions $F$ and $G$ for each new $(x_u, y_u^{-})$ pair results in linear computational growth with the negative sample size.

To address these computational challenges, NAPO directly shares the precomputed log probabilities (\ie values of $H_{\theta}$) across sequences within each batch, as illustrated in Figure~\ref{method}(b).
This strategy effectively expands negative coverage without incurring the computational overhead of either computing all logits-negative combinations or recalculating functions $F$ and $G$ for shared negatives.
However, the validity of such sharing critically depends on the prompt $x$, as the same item may have different relevance scores under different user contexts.
For example, the $H_{\theta}$ value computed for a science fiction enthusiast may not be applicable to a user preferring romantic comedies. 

To ensure contextual validity, we introduce a similarity-guided filtering mechanism based on a pre-trained lightweight sequential recommendation (\eg SASRec \cite{DBLP:conf/icdm/KangM18}).
Specifically, following \cite{DBLP:journals/corr/abs-2312-02445}, we extract the sequence representation $\text{SR-EMB}(s)$ from the sequential recommender to capture the user's underlying preference \cite{DBLP:journals/corr/abs-2408-10159}, based on which, we calculate the similarity between two sequences $s_u$ and $s_v$ as:
\begin{equation}
\text{sim}(s_u, s_v) = < \text{SR-EMB}(s_u), \text{SR-EMB}(s_v)>,
\end{equation}
where $<\cdot,\cdot>$ denotes inner product.
The set of sequences eligible for sharing $H_{\theta}$ values with sequences $s_u$ consists of the Top-$K$ most similar sequences to $s_u$,
% \begin{equation}
%     \mathcal{T}_u = \{ s_v \mid s_v \in \text{Top-}K(\text{sim}(s_u, :)) \wedge s_v \neq s_u \}.
% \end{equation}
with the number of similar sequences $K$ determined by: $K = \left\lfloor \left( b_p  - 1 \right) \times \rho \right\rfloor,$ where $b_p$ denotes the batch size and $\rho$ controls the proportion of similar sequences to be shared. 

Through this similarity-guided sharing strategy, we effectively expand the negative sample coverage to $N_{neg} = n_{neg} + K  \times n_{neg}$ without additional forward passes through the LLM.
Here, $n_{neg} = |\hat{\mathcal{E}}|$ is the number of negative items sampled for each sequence.

\subsection{Dynamic $\gamma$ Adjustment}
\label{ssec:dynamic_gamma}
While expanding negative sample coverage through in-batch sharing enhances the model's discriminative ability, treating all negative signals equally may not be optimal.
The impact of different negative samples on preference optimization can vary significantly \cite{DBLP:journals/corr/abs-2410-10148,DBLP:journals/corr/abs-2407-08639}.
This heterogeneity stems from both the inherent characteristics of items and the user's preference patterns.
To effectively leverage these varying negative signals, we propose a dynamic adjustment mechanism for the reward margin $\gamma$ that adapts to the confidence level of negative samples.

In the NAPO objective (\cf Equation~\eqref{eq:napo_loss}), $\gamma$ serves as a reward margin that adjusts the difference between positive and negative items.
While a fixed $\gamma_0$ as in SimPO offers simplicity, it fails to capture the different impacts of negative samples with varying levels of certainty.
To address this limitation, we introduce a confidence-based adjustment mechanism that modifies $\gamma$ based on the reliability of unobserved interactions.
This strategy ensures that high-confidence negatives --- those more likely to be true negatives --- exert stronger influence by enlarging the value of $\gamma$, while reducing the impact of uncertain cases in an opposite way.
We utilize the same auxiliary sequential recommendation in Section \ref{ssec:in-batch} to compute a confidence score for each $(s_u, y_u^{-})$ pair, where $y_u^{-} \in \mathcal{E}_u$.
Formally, we compute the confidence score between sequence $s_u$ and negative item $y_u^{-}$ as:
\begin{equation}
    \text{conf}(s_u, y_u^{-}) =  \frac{1}{2}\left(1-\text{SR-SCORE}(s_u,y_u^{-})\right),
\end{equation}
where $\text{SR-SCORE}(s_u,y_u^{-})$ represents the rating score, measuring the relevance between the sequence $s_u$ and the negative item $y_u^{-}$. The score ranges from $\left[-1, 1\right]$ and is normalized to $\left[0, 1\right]$ to compute the confidence score.
A higher relevance indicates lower confidence in the negative sample's validity, as such a sample lies closer to the decision boundary \cite{DBLP:journals/ir/ChapelleK10,DBLP:books/daglib/0097035,platt1999probabilistic}.
Conversely, a lower relevance reflects higher confidence in the sample's reliability.

To incorporate this confidence information, we draw inspiration from \cite{DBLP:journals/corr/abs-2407-08639}, and propose a batch-level dynamic adjustment strategy:
\begin{equation}
    \gamma_{\text{batch}} = \left[1 + \alpha\left(\mathbb{E}_{u \sim \mathcal{B},y_u^- \sim \mathcal{E}_u}\left(\text{conf}(s_u, y_u^{-})\right) - R_0\right)\right]\gamma_0,
\end{equation}
where $\gamma_0$ is the initial margin value, and $\alpha$ is an adjustment coefficient that controls the sensitivity of the margin to confidence fluctuations. The term $\mathbb{E}_{u \sim \mathcal{B},y_u^- \sim \mathcal{E}_u}\left(\text{conf}(s_u, y_u^{-})\right)$ computes the average confidence score of negative samples within the batch $\mathcal{B}$, while $R_0$
  is a baseline confidence threshold that serves as a reference for margin adjustment. When the batch’s average confidence exceeds $R_0$, the margin $\gamma_{\text{batch}}$ is increased, and when it is lower, the margin is decreased.
To maintain stability,  $R_0$ is adaptively updated using a momentum-based moving average:
\begin{equation}
    R_0 \leftarrow mR_0 + (1-m)\mathbb{E}_{u \sim \mathcal{B},y_u^- \sim \mathcal{E}_u}\left(\text{conf}(s_u, y_u^{-})\right),
\end{equation}
with momentum coefficient $m \in [0,1)$. A higher $m$ makes $R_0$
more stable but slower to adapt, while a lower $m$ makes it more reactive to recent confidence changes. This batch-level strategy not only maintains computational efficiency by computing adjustments at the batch level, but also offers more stable training by avoiding the local instabilities (\eg noisy negative signals) that might arise from per-instance update \cite{DBLP:journals/corr/abs-2407-08639}.

%% file: chapters/4_exp.tex
\section{Experiment}
We evaluate the performance of NAPO through experiments on three widely used benchmark datasets: LastFM \cite{DBLP:conf/recsys/2011hetrec}, Goodreads\footnote{\url{https://www.goodreads.com.}}, and Steam \cite{DBLP:conf/icdm/KangM18}. We organize the sequences chronologically to preserve the temporal nature of interactions. We select $9$ state-of-the-art method: {Traditional methods}: GRU4Rec \cite{DBLP:journals/corr/HidasiKBT15}, Caser \cite{DBLP:conf/wsdm/TangW18}, and SASRec \cite{DBLP:conf/icdm/KangM18}. {LLM-based methods}: LLaMA2 \cite{DBLP:journals/corr/abs-2307-09288}, ChatRec \cite{gao2023chat}, MoRec \cite{DBLP:conf/sigir/YuanYSLFYPN23}, TALLRec \cite{DBLP:conf/recsys/BaoZZWF023}, LLaRA \cite{DBLP:journals/corr/abs-2312-02445}, and S-DPO \cite{DBLP:journals/corr/abs-2406-09215}. Our objective is to recommend items from a candidate set to users. Due to LLMs' challenges in generating ranked lists, position-aware metrics like NDCG are unsuitable \cite{DBLP:journals/corr/abs-2406-09215}. Instead, we employ HitRatio@1 for top-item relevance and ValidRatio for effective response proportion. Comprehensive details on datasets, baselines, and evaluation metrics can be found in Appendix~\ref{ref:exp}. We focus on the following research questions:
\textbf{RQ1:} How does NAPO perform compared to traditional and LLM-based sequential recommendation models? \textbf{RQ2:} What is the impact of our design (\ie in-batch negative sharing strategy and dynamic impact
assessing mechanism) on recommendation performance? 
 \textbf{RQ3:} How does NAPO perform in reducing popularity bias?

\begin{table*}[t]
\begin{center}
\small
\caption{Experimental results on three real-world datasets. The term "Rel.Ipv" refers to the relative improvement of NAPO compared with the baselines. The best results are marked in bold.}
\label{exp:main}
\resizebox{\textwidth}{!}{
\begin{tabular}{ccccccccccc}
\toprule
\textbf{\multirow{2}{*}{Category}} & \textbf{\multirow{2}{*}{Method}}
  & \multicolumn{3}{c}{\textbf{Goodreads}}   & \multicolumn{3}{c}{\textbf{LastFM}}      & \multicolumn{3}{c}{\textbf{Steam}}  \\ \cmidrule(lr){3-5}  \cmidrule(lr){6-8} \cmidrule(lr){9-11} 
&& \textbf{HitRatio@1}   & \textbf{ValidRatio} & \textbf{Rel.Ipv}   & \textbf{HitRatio@1}   & \textbf{ValidRatio} & \textbf{Rel.Ipv}   & \textbf{HitRatio@1}   & \textbf{ValidRatio} & \textbf{Rel.Ipv}   \\
\midrule
\textbf{\multirow{3}{*}{Traditional}} & \textbf{GRU4Rec} \cite{DBLP:journals/corr/HidasiKBT15} & 0.3867 & 1.0000     & 92.86\%   & 0.2616 & 1.0000     & 166.09\%  & 0.4168 & 1.0000     & 97.21\%  \\
                             & \textbf{Caser} \cite{DBLP:conf/wsdm/TangW18}   & 0.4174 & 1.0000     & 78.67\%   & 0.2233 & 1.0000     & 211.73\%  & 0.4368 & 1.0000     & 88.18\%  \\
                             & \textbf{SASRec} \cite{DBLP:conf/icdm/KangM18}  & 0.3581 & 1.0000     & 108.26\%   & 0.2233 & 1.0000     & 211.73\%  & 0.4010 & 1.0000     & 104.98\%  \\ \midrule
\textbf{\multirow{6}{*}{LLM-based}}    & \textbf{LLaMA2} \cite{DBLP:journals/corr/abs-2307-09288}  & 0.0233 & 0.3845     & 3100.85\% & 0.0246 & 0.3443     & 2729.67\% & 0.0135 & 0.1653     & 5988.88\% \\
                             & \textbf{ChatRec} \cite{gao2023chat}  & 0.3306 & 1.0000    & 125.58\%    & 0.3770 & 1.0000    & 84.64\% & 0.3626 &   0.9798    & 126.69\% \\
                             & \textbf{MoRec} \cite{DBLP:conf/sigir/YuanYSLFYPN23} & 0.2877 & 1.0000     & 159.22\%  & 0.1652 & 1.0000     & 321.36\%  & 0.3911 & 1.0000     & 110.17\%  \\
                             & \textbf{TALLRec} \cite{DBLP:conf/recsys/BaoZZWF023}   & 0.4983 & 0.9573     & 49.66\%   & 0.4180 & 0.9836     & 66.53\%   & 0.4637 & 0.9840     & 77.26\%   \\
                             & \textbf{LLaRA} \cite{DBLP:journals/corr/abs-2312-02445}   & 0.5292 & 0.9950     & 40.92\%   & 0.4508 & 0.9918     & 54.41\%   & 0.4949 & 0.9975     & 66.09\%   \\
                             & \textbf{S-DPO} \cite{DBLP:journals/corr/abs-2406-09215}   & 0.6661 & 0.9950     & 11.96\%          & 0.6477 & 0.9980     &7.47\%           & 0.6703 & 0.9881     &  22.63\%      \\ \midrule
\rowcolor{gray!20} 
\textbf{Ours}                         & \textbf{NAPO}  &\textbf{0.7458}        &0.9933            & -          &     \textbf{0.6961}   &  0.9979          &   -    &      \textbf{0.8220}   & 0.9932           & -          \\ \bottomrule
\end{tabular}}
\vspace{-10pt}
\end{center}
\end{table*}

\subsection{Overall Performance Comparison (RQ1)}
The experimental results on three datasets are presented in Table~\ref{exp:main}. From these results, we can obtain the following observations:  \textbf{(1) LLM4Rec methods demonstrate significant potential compared to traditional recommendation approaches.} LLM-based methods significantly improve HitRatio@1 over traditional baselines due to their extensive knowledge and advanced reasoning. MoRec underperforms as it fails to leverage LLMs' reasoning effectively. \textbf{(2) Vanilla LLMs struggle with recommendation tasks.}
    For instance, LLaMA2 achieves low ValidRatios of $0.3845$, $0.3443$, and $0.1653$ on Goodreads, LastFM, and Steam due to a lack of domain-specific guidance. In contrast, SFT-based methods improve ValidRatio by fine-tuning on interaction data. S-DPO and NAPO retain LLMs' core capabilities, achieving high ValidRatio through preference alignment. \textbf{(3) The preference alignment stage further enhances the performance of LLM4Rec.} S-DPO and NAPO outperform other LLM-based methods by leveraging multi-negative samples. NAPO further surpasses S-DPO with a 14.02\% average increase in HitRatio@1, demonstrating its superior handling of negative samples. \textbf{(4) NAPO consistently outperforms all baselines.} It achieves significant HitRatio@1 gains over the second-best baseline, with improvements of 11.96\%, 7.47\%, and 22.63\% on Goodreads, LastFM, and Steam, respectively. These results highlight NAPO's effective in-batch negative sharing and dynamic impact assessment for efficient negative sample utilization.

\subsection{Impact of Two Key Designs in NAPO (RQ2)}
\begin{figure}[t]
  \centering
  % 左侧图像，占单栏一半宽度
  \begin{minipage}[b]{0.48\textwidth}
    \centering
    \begin{minipage}[b]{0.48\textwidth}
      \centering\includegraphics[width=\textwidth]{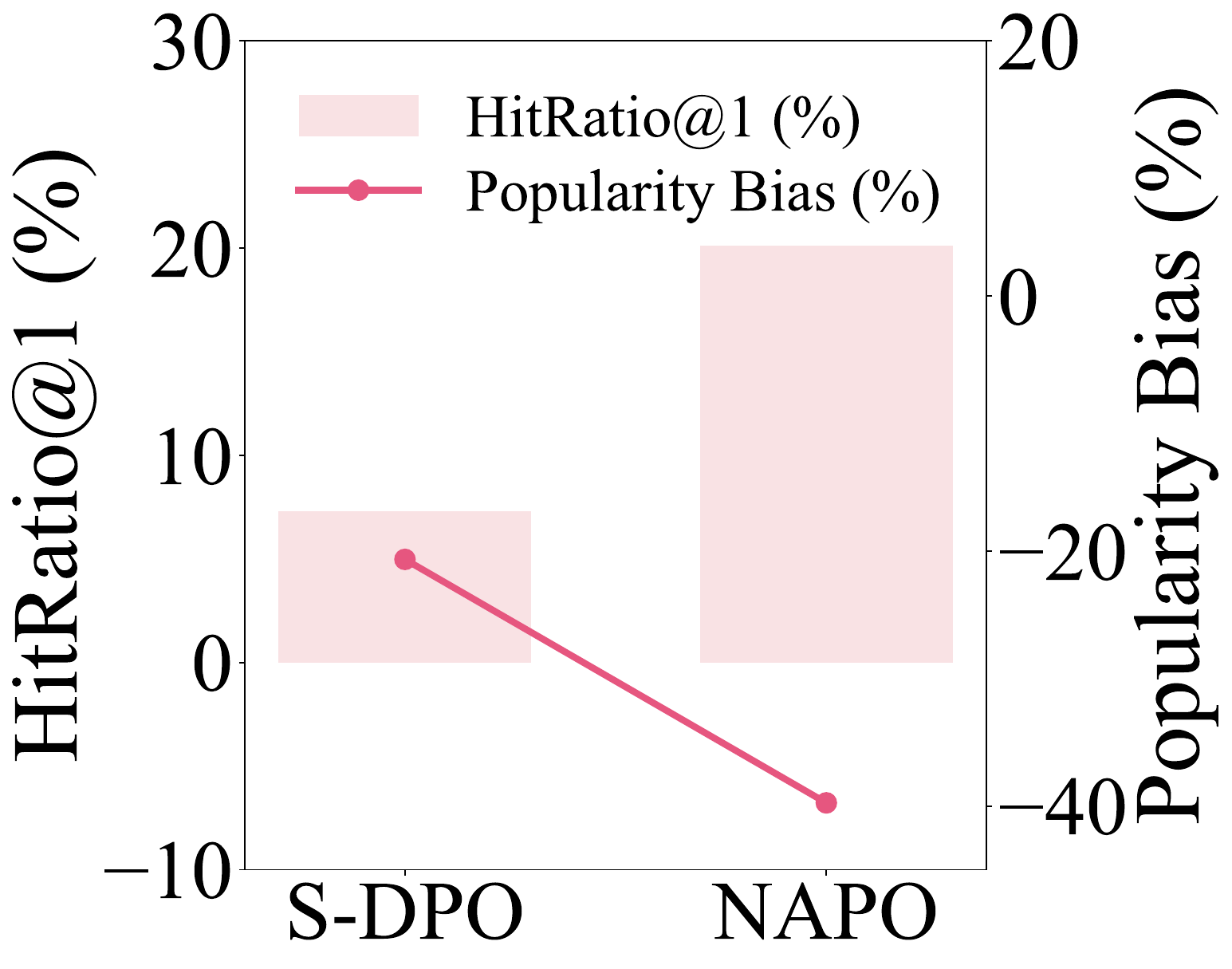}
      \vspace{-8pt}
      \subcaption{Results on Goodreads.}
      \label{bias1}
    \end{minipage}
    \hfill
    \begin{minipage}[b]{0.48\textwidth}
      \centering
\includegraphics[width=\textwidth]{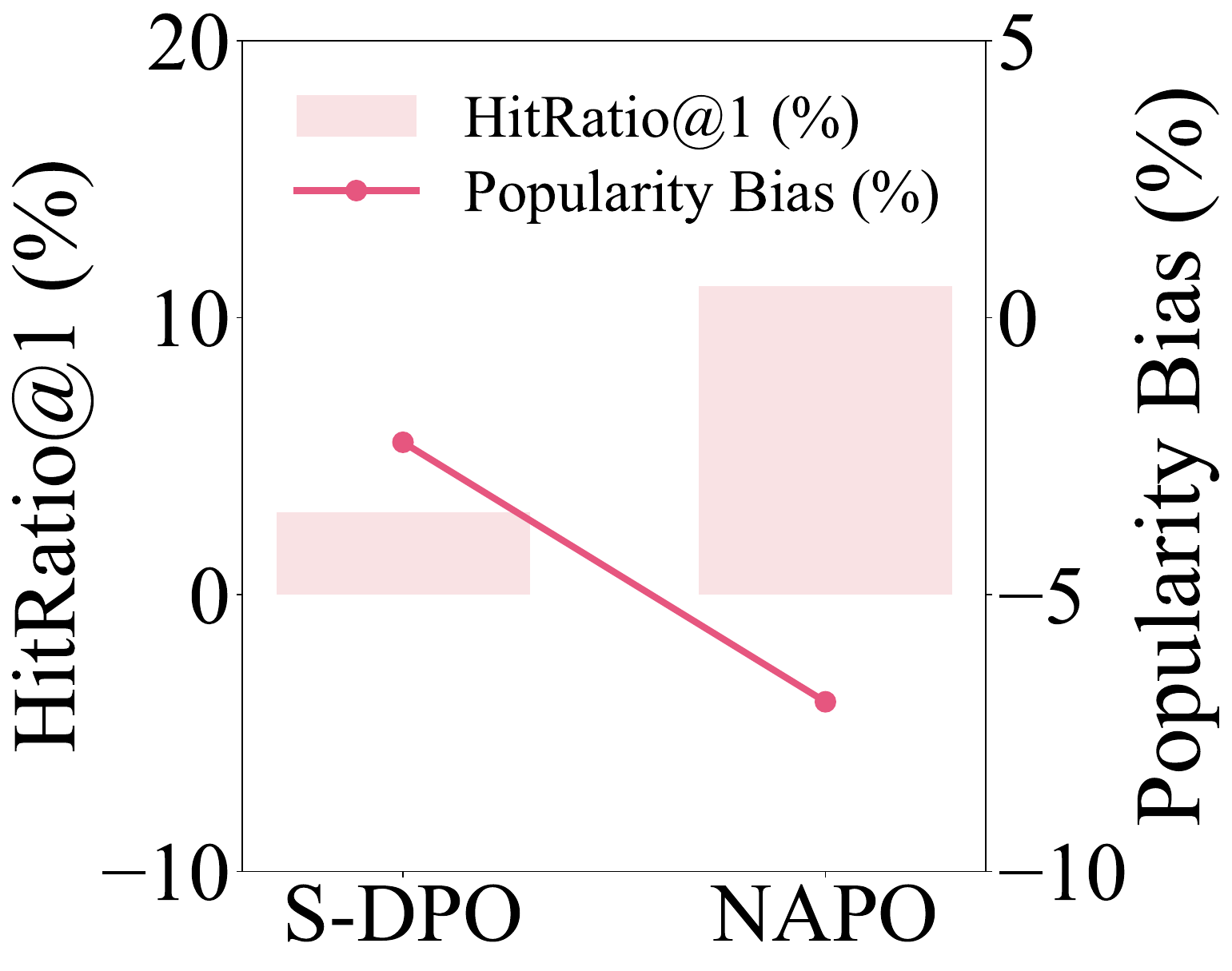}
      \vspace{-8pt}
      \subcaption{Results on LastFM.}
      \label{bias2}
    \end{minipage}
  \end{minipage}%
  \hfill
  \begin{minipage}[b]{0.5\textwidth}
\centering
\includegraphics[width=\textwidth]{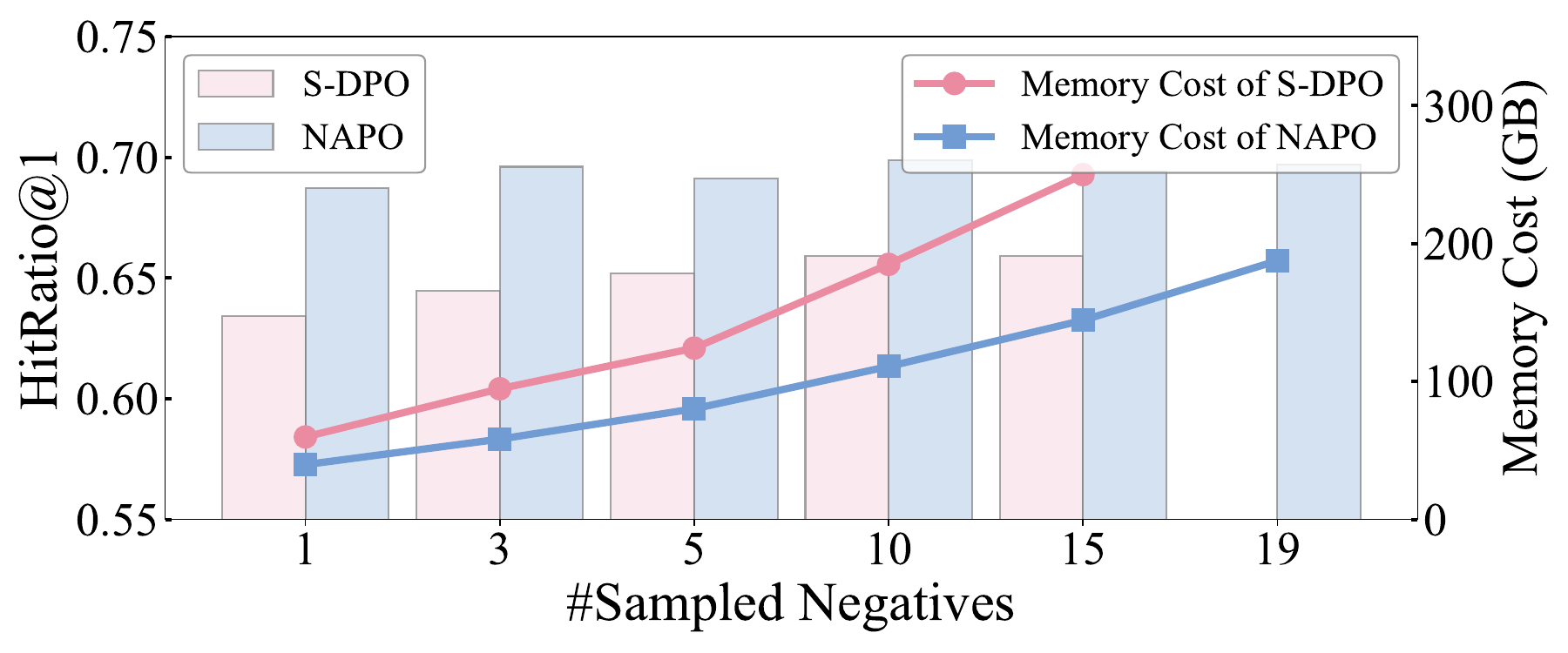}
\subcaption{Results for varying negative samples.}
    \label{number}
  \end{minipage}
  \end{figure}

We conduct ablation experiments on NAPO's key components across two datasets to assess their contributions. 
The components studied include: length normalization (w/ Length Norm), in-batch {N}egative {S}haring (w/o NS), Dynamic $\gamma$ Adjustment (w/o Dynamic $\gamma$), in-batch Negative Sharing combined with Dynamic $\gamma$ Adjustment (w/o NS \& Dynamic $\gamma$), and in-batch Negative Sharing combined with hyper-parameter $\gamma$ (w/o NS \& $\gamma$). 
Additionally, we perform experiments on S-DPO \cite{DBLP:journals/corr/abs-2406-09215} to study the effectiveness of our in-batch Negative Sharing (S-DPO + NS) and further adapt the S-DPO framework to the SimPO \cite{DBLP:journals/corr/abs-2405-14734} setting (SimS-DPO). The results of these ablation studies are presented in Table~\ref{ablation}. 
\begin{wraptable}{l}{9cm}
\setlength{\tabcolsep}{3pt}
\caption{Ablation studies on the Goodreads and LastFM datasets. "HR@1" is an abbreviation for HitRatio@1.}
\label{ablation}
\vspace{-5pt}
\small
\begin{tabular}{lcccc}
\toprule
\multicolumn{1}{c}{\multirow{2}{*}{\textbf{Method}}} & \multicolumn{2}{c}{\textbf{Goodreads}} & \multicolumn{2}{c}{\textbf{LastFM}} \\ \cmidrule(lr){2-3} \cmidrule(lr){4-5}
& \textbf{HR@1} & \textbf{ValidRatio} & \textbf{HR@1} & \textbf{ValidRatio} \\ \midrule
\textbf{S-DPO} & 0.6661 & 0.9950 & 0.6448 & 0.9980\\ 
\textbf{SimS-DPO} & 0.6727 & 0.9518 & 0.6501 & 0.9927     \\
\textbf{S-DPO + NS} & 0.7408 & 0.9950 & 0.6857 & 0.9980 \\
\rowcolor{gray!20} 
\textbf{NAPO} & \textbf{0.7458} & 0.9980 & \textbf{0.6961} & 0.9934 \\
\quad w/ Length Norm & 0.7392 & 0.9950 & 0.6880 & 0.9980\\
\quad w/o NS & 0.6727 & 0.9668 & 0.6549 & 0.9920\\ 
\quad w/o Dynamic $\gamma$ & 0.7241 & 0.9933 & 0.6797 & 0.9984\\
\quad w/o NS \& $\gamma$ & 0.6727 & 0.9518 & 0.6501 & 0.9927\\ 
\quad w/o NS \& Dynamic $\gamma$ & 0.6694 & 0.9501 & 0.6484 & 0.9923\\
\bottomrule
\end{tabular}
\vspace{-10pt}
\end{wraptable}
From these experimental results, we can obtain the following observations: \textbf{(1) In-batch negative sharing boosts performance.} Removing this strategy reduces NAPO's performance (from $0.7458$ to $0.6727$ on Goodreads and $0.6961$ to $0.6549$ on LastFM), while adding it to S-DPO improves results (from $0.6661$ to $0.7408$ on Goodreads and $0.6448$ to $0.6857$ on LastFM), emphasizing the impact of effective negative integration. \textbf{(2) Dynamic $\gamma$ adjustment enhances NAPO.} Both NAPO variants --- with and without in-batch negative sharing (NS) --- perform better with dynamic $\gamma$, while constant $\gamma$ leads to suboptimal results, highlighting the need for adaptive negative sample handling. \textbf{(3) Length normalization may hinder NAPO.} NAPO's performance drops with length normalization (from $0.7458$ to $0.7392$ on Goodreads and $0.6961$ to $0.6880$ on LastFM), likely due to structural similarity in the response to LLM for recommendation task.

% \begin{figure}[t]
%     \centering
%  \includegraphics[width=0.35\textwidth]{figures/number_with_memory_cost_separated_legends.pdf}
%     \vspace{-10pt}
%     \caption{Experimental results of HitRatio@1 and Memory Cost (GB) for S-DPO and NAPO across varying numbers of sampled negatives.}
%     \label{fig:num_neg_samples}
%     \vspace{-15pt}
% \end{figure}

\subsection{Analysis of Performance in Reducing Popularity Bias (RQ3)}
In this section, we examine the impact of NAPO and S-DPO on reducing popularity bias. Following prior work \cite{DBLP:journals/corr/abs-2410-12519}, we define popularity bias as the deviation between the popularity of a recommended item and the average popularity of items within a user's historical sequence. We evaluate the popularity bias by computing the expectation $\mathbb{E}[\text{bias}_{pop}(u, i_t)]$, with detailed calculation method provided in Appendix~\ref{ref:bias}. Specifically, we calculate the relative improvement in {HitRatio@1} and \( \mathbb{E}[\text{bias}_{pop}(u, i_t)] \) for both S-DPO and NAPO, expressed as a percentage compared to the SFT model. A lower percentage of \( \mathbb{E}[\text{bias}_{pop}(u, i_t)] \) indicates a greater reduction in popularity bias, while a higher percentage of {HitRatio@1} indicates a greater performance improvement. The recommendation performance and popularity bias comparisons for S-DPO and NAPO are shown in Figure~\ref{bias1} and Figure~\ref{bias2}. Both NAPO and S-DPO enhance HitRatio\@1 and mitigate popularity bias compared to the SFT model. Notably, NAPO outperforms S-DPO, achieving higher recommendation accuracy and stronger bias reduction.

\subsection{Study of NAPO}\label{parameter}
In this section, we analyze the impact of negative sample quantity to better understand NAPO. For hyperparameter experiments, please refer to the Appendix~\ref{ref:hyper}. Here, $n_{neg}$ denotes the number of sampled negatives before in-batch negative sharing. Figure~\ref{number} compares the performance and GPU memory cost of S-DPO and NAPO under varying $n_{neg}$. As $n_{neg}$ increases, S-DPO's performance steadily improves, while NAPO's fluctuates. This indicates that excessive negative samples may saturate performance, as seen when $n_{neg}=3$ for NAPO, where the equivalent number of negatives reaches $33$ without further gains.
Despite this, NAPO consistently outperforms S-DPO. Even when $n_{neg} = 15$ for S-DPO and $n_{neg} = 1$ for NAPO, NAPO still maintains superior performance.

In terms of GPU memory, both methods consume more memory as $n_{neg}$ increases. However, S-DPO has higher memory usage due to the reference model $\pi_\text{ref}$, while NAPO remains efficient, leveraging in-batch negative sharing without extra overhead. Furthermore, due to GPU memory limitations, S-DPO can only accommodate up to $15$ negative samples, while NAPO can handle up to $19$ negative samples (equivalent to $209$ negative samples in optimization). Experimental results on time efficiency can be found in Appendix~\ref{ref:time}.

%% file: chapters/6_conclusion.tex
\section{Conclusion}
We introduce Negative-Aware Preference Optimization (NAPO), a framework designed to enhance the utilization of negative samples in LLM-based recommenders. It addresses two challenges: scalable integration of negatives and adaptive impact adjustment. Through an in-batch negative sharing strategy, NAPO effectively broadens the negative sample space without increasing computational overhead. It further employs dynamic reward margin adjustment, prioritizing high-confidence negatives to improve model updates. Experimental results across multiple datasets confirm NAPO's enhanced effectiveness, demonstrating improved handling of negative samples and strengthening recommendation performance. Our solution offers a scalable, adaptive approach, pushing forward advancements in recommendation methodologies.

Despite the advantages of our approach, there are still some challenges. Specifically, as the number of sampled negatives increases, computational overhead grows significantly. Additionally, compared to S-DPO, our method introduces more hyperparameters, requiring more time for tuning.

%% file: chapters/7_appendix.tex
\newpage
\appendix
\section{Experimental Details} \label{ref:exp}
\subsection{Dataset}
We follow the same data preprocessing approach as S-DPO \cite{DBLP:journals/corr/abs-2406-09215} for fair comparisons. Specifically, the datasets are split into training, validation, and test sets in an $8:1:1$ ratio. For the Goodreads, we filter out users and items with fewer than $20$ interactions. The statistics of these datasets are summarized in Table~\ref{dataset}.
\begin{table}[h]
\caption{Statistics of Goodreads and LastFM datasets.}
\label{dataset}
\begin{center}
\setlength{\tabcolsep}{7pt}
\small
\begin{tabular}{crrr}
\toprule
Dataset        & Goodreads & LastFM & Steam \\ \midrule
\#Sequence     & 6,031     & 1,220  & 11,938 \\
\#Items        & 4,500     & 4,606  &  3,581 \\
\#Interactions & 220,100   & 73,510 & 274,726 \\ \bottomrule
\end{tabular}
\end{center}
\end{table}

\subsection{\textbf{Baselines}} We select $9$ state-of-the-art methods, spanning both traditional and LLM-based models, for a comprehensive comparison.

\noindent \textbf{- Traditional methods}:
% (1) {GRU4Rec \cite{DBLP:journals/corr/HidasiKBT15}} leverages GRU (Gated Recurrent Unit) architecture to model sequential user behavior.
% (2) {Caser \cite{DBLP:conf/wsdm/TangW18}} utilizes convolutional filters to model the sequential relationships in user behavior.
% (3) {SASRec \cite{DBLP:conf/icdm/KangM18}} uses a self-attention mechanism to capture both long-term dependencies and recent user behavior.
% \noindent \textbf{- LLM-based methods}:
%  (1) {LLaMA2 \cite{DBLP:journals/corr/abs-2307-09288}:}  We leverage LLaMA2-7B to generate recommendation results through direct prompting without additional fine-tuning or specialized training.
% (2) {ChatRec \cite{gao2023chat}}  integrates LLMs with traditional recommendation systems by converting user profiles and interaction histories into prompts.
% (3) {MoRec \cite{DBLP:conf/sigir/YuanYSLFYPN23}} replaces item ID features with modality-specific representations, utilizing BERT \cite{DBLP:journals/corr/abs-1810-04805} to encode text and SASRec to model sequential user behavior.
% (4) {TALLRec \cite{DBLP:conf/recsys/BaoZZWF023}} proposes an efficient and effective tuning framework that aligns LLMs with recommendation tasks through instruction tuning.
% (5) {LLaRA \cite{DBLP:journals/corr/abs-2312-02445}} integrates traditional sequential recommenders with LLMs through hybrid prompting and curriculum learning.
% (6) {S-DPO \cite{DBLP:journals/corr/abs-2406-09215}} improves recommendation performance by integrating multiple negative samples and adapting the Plackett-Luce preference model for better ranking.
(1) GRU4Rec \cite{DBLP:journals/corr/HidasiKBT15}: Utilizes GRU to model sequential user behavior.  (2) Caser \cite{DBLP:conf/wsdm/TangW18}: Employs convolutional filters to capture sequential user behavior.  (3) SASRec \cite{DBLP:conf/icdm/KangM18}: Uses self-attention to model long-term dependencies and recent behavior.

\noindent \textbf{- LLM-based methods}:
(1) LLaMA2 \cite{DBLP:journals/corr/abs-2307-09288}: Generates recommendations via direct prompting without fine-tuning.  (2) ChatRec \cite{gao2023chat}: Converts user profiles and histories into prompts, integrating LLMs with traditional systems.  (3) MoRec \cite{DBLP:conf/sigir/YuanYSLFYPN23}: Uses BERT to encode text and SASRec to model sequences, replacing item IDs with modality-specific representations.  (4) TALLRec \cite{DBLP:conf/recsys/BaoZZWF023}: Aligns LLMs with recommendation tasks via instruction tuning. (5) LLaRA \cite{DBLP:journals/corr/abs-2312-02445}: Combines traditional sequential recommenders with LLMs through hybrid prompting and curriculum learning.  (6) S-DPO \cite{DBLP:journals/corr/abs-2406-09215}: Enhances performance by integrating multiple negative samples and refining ranking with the Plackett-Luce model. 

\subsection{\textbf{Implementation Details}}
All approaches are implemented using Python 3.9.7, PyTorch 2.2.2, and Transformers 4.43.3, running on 4 NVIDIA A100 GPUs. We select LLaMA2-7B \cite{DBLP:journals/corr/abs-2307-09288} as the LLM backbone. 
% Traditional methods are optimized using the Adam optimizer with a learning rate of $0.001$ and a batch size of $256$. L2 regularization is applied to all models, with the coefficient empirically tuned from the range [1e-3, 1e-4, 1e-5, 1e-6, 1e-7]. 
The LLM-based methods are trained for a maximum of $5$ epochs with a batch size of 128.
% As suggested by \cite{DBLP:journals/corr/abs-2312-02445}, we adopt a random sampling strategy for prompts across multiple formats during training and evaluation, enabling flexibility and broad generalization. 
Specifically, for S-DPO and NAPO, preference training is conducted for an additional $3$ epochs with a batch size of $128$ and a learning rate of 1e-5. We configure $\beta = 1$ and use $3$ negative samples for S-DPO. The hyperparameters in NAPO are set as $\gamma_{0} = 1.0$, $\rho=0.7$, and $\alpha = 0.3$. 
% For distributed training, we set $ nproc\_per\_node = 4$, $batch\_size\_per\_node = 4$, and $gradient\_accumulation\_steps = 8$. Further analysis of the hyperparameters can be found in Section~\ref{parameter}. 
We follow the same experimental setup as S-DPO \cite{DBLP:journals/corr/abs-2406-09215} and directly use the corresponding results for LastFM and Goodreads from its main table. The results for the Steam dataset are referenced from the Table 2 of LLaRA \cite{DBLP:journals/corr/abs-2312-02445}, as their experimental setups are consistent.

\section{\textbf{Analysis of Different Dynamic $\gamma$ Adjustment Mechanism}} \label{ref:alpha}
To validate the effectiveness of our dynamic $\gamma$ adjustment strategy, we conduct experiments comparing it with a personalized $\gamma$ adjustment strategy \cite{DBLP:journals/corr/abs-2410-10148} across two datasets. Specifically, $\alpha$-DPO \cite{DBLP:journals/corr/abs-2410-10148} utilizes additional information from $\pi_\text{ref}$ to automatically adjust the reward margin $\gamma$. We compare NAPO with SimPO, SFT, and several other variants incorporating $\alpha$-DPO's reward margin adjustment strategy. Additionally, we set $n_{neg} = 1$ across all methods for fair comparison.  

% In particular, our NAPO method, which originally calculates $\gamma_{\text{NAPO}}$ using the proposed dynamic $\gamma$ adjustment strategy, is modified in two variants. 

\begin{figure}[t]
  \centering
  \begin{minipage}[b]{0.3\textwidth}
    \centering
    \includegraphics[width=\textwidth]{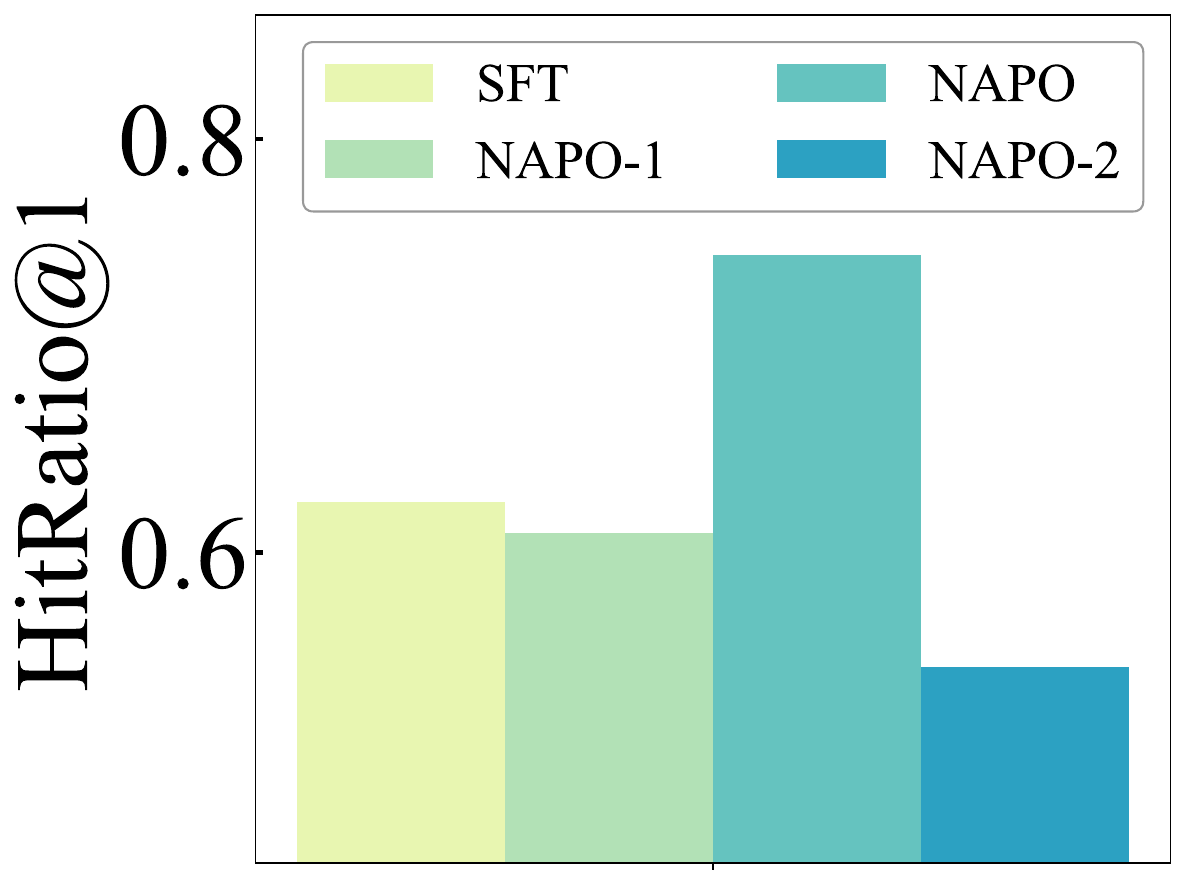}
    \subcaption{Results on Goodreads.}
  \end{minipage}
  \begin{minipage}[b]{0.3\textwidth}
    \centering
    \includegraphics[width=\textwidth]{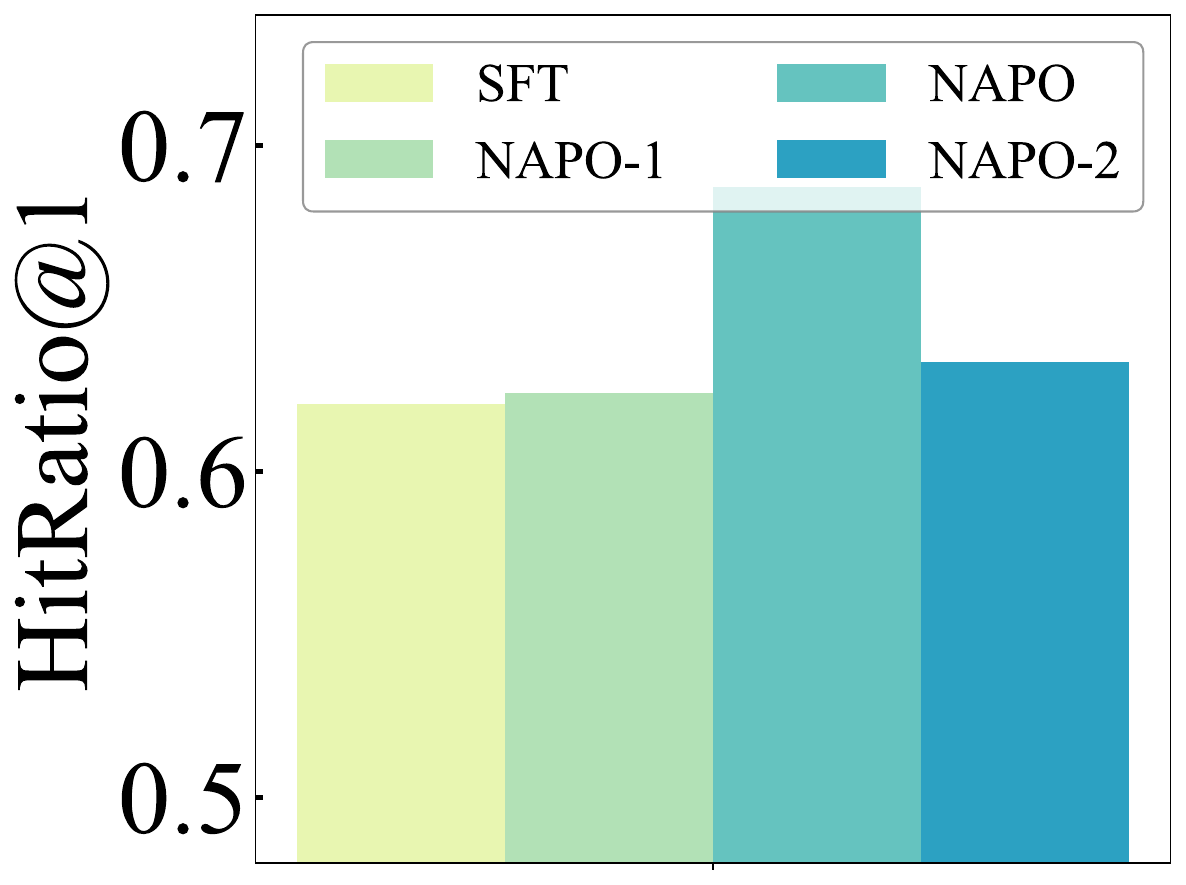}
    \subcaption{Results on LastFM.}
    \end{minipage}
  \caption{Experimental result for SFT, NAPO, and its variants.}
  \label{fig:compare_with_alpha_dpo}
\end{figure}

For NAPO-1, we replace $\gamma_{\text{NAPO}}$ with $\gamma_{\alpha_{\text{DPO}}}$, which is calculated using the $\alpha$-DPO strategy. For NAPO-2, we combine both adjustment strategies, where $\gamma$ is calculated as $\gamma = \frac{\gamma_{\text{NAPO}} + \gamma_{\alpha{\text{-DPO}}}}{2}$. The experimental results are shown in Figure~\ref{fig:compare_with_alpha_dpo}. We observe that the standard NAPO method consistently outperforms all other methods. Specifically, when taking the $\alpha$-DPO strategy or combining both the dynamic $\gamma$ adjustment and $\alpha$-DPO strategies, performance declines on the Goodreads dataset, with results even falling below those of the SFT model. 
% On the LastFM dataset, while there is a slight improvement compared to SFT, the gains are marginal and can be considered negligible. 
Overall, these findings suggest that the $\alpha$-DPO reward margin adjustment strategy may not be well-suited for recommendation tasks.

\section{Hyperparameter Analysis}\label{ref:hyper}
\subsection{\textbf{Study on Values of $\rho$}}
The term $\rho$ denotes the proportion of similar sequences, guiding negative sample selection by filtering out irrelevant ones. Figure~\ref{ff1} shows that performance improves as $\rho$ increases due to a larger pool of shared negatives. However, excessively high $\rho$ introduces noise, leading to performance decline.
% Based on these observations, we set $\rho$ to $0.7$.

% \subsubsection{\textbf{Study on Values of $\gamma_0$.}}
% The initial value of $\gamma$, represented by $\gamma_0$, is shown to impact performance, as illustrated in Figure~\ref{ff2}. We observe that when $\gamma_0 = 0$, the performance is suboptimal. Introducing the reward margin leads to enhanced performance, and as $\gamma_0$ increases, the recommendation performance further improves. However, higher values of $\gamma_0$ do not always result in better performance. Taking this into account, we set $\gamma_0$ to $1.0$.

\subsection{\textbf{Study on Values of $\alpha$}}
\begin{figure}[h]
  \centering
    \begin{minipage}[t]{0.40\textwidth}
        \centering
        \includegraphics[width=\textwidth]{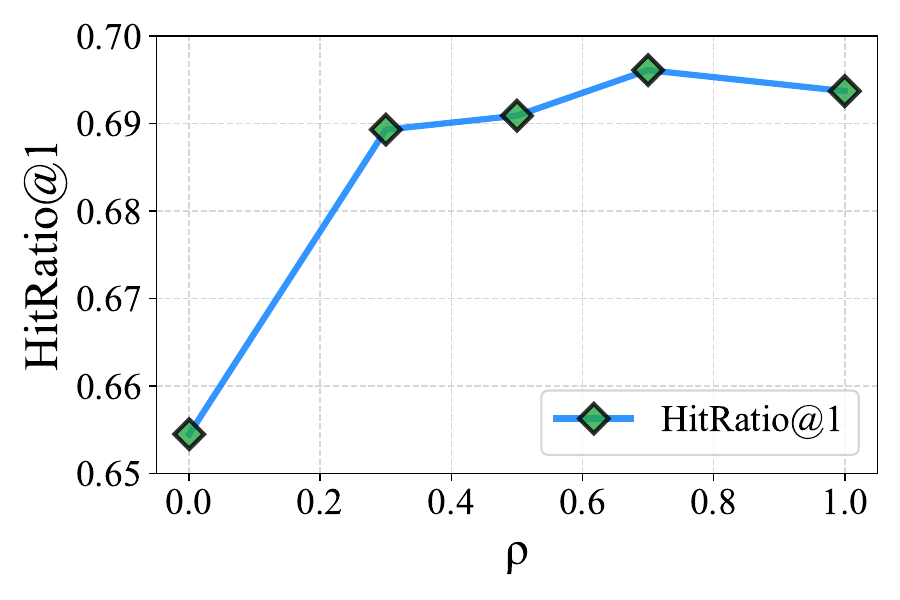}
        \vspace{-15pt}
        \subcaption{Study of $\rho$ on HitRatio@1.}
        \label{ff1}
    \end{minipage}
    \begin{minipage}[t]{0.40\textwidth}
        \centering
        \includegraphics[width=\textwidth]{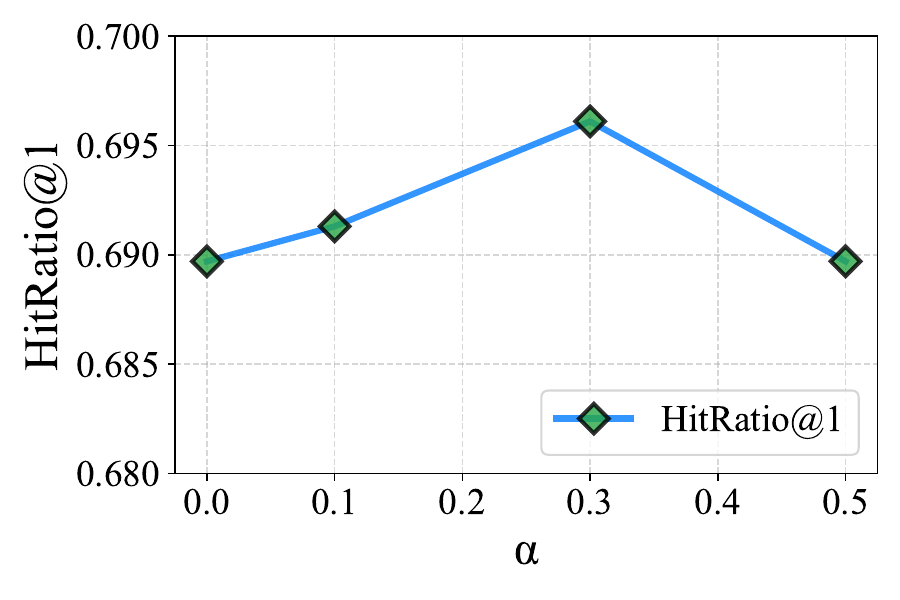}
        \subcaption{Study of $\alpha$ on HitRatio@1.}
        \label{ff3}
    \end{minipage}
  \caption{Studies on values of $\rho$ and $\alpha$ of NAPO on LastFM. (\ref{ff1}): Experimental results with varying values of $\rho$. (\ref{ff3}): Experimental results with varying values of $\alpha$.}
  \label{fig:both_images}
\end{figure}

The parameter $\alpha$ adjusts the reward margin for negative samples. Figure~\ref{ff3} shows that performance improves with increasing $\alpha$ due to stronger margin adjustments but declines beyond a certain point, likely due to excessive adjustments.
% Based on these findings, we set $\alpha$ to $0.3$ for optimal performance.

\section{Evaluation of Popularity Bias}\label{ref:bias}
Specifically, the popularity bias is expressed as: 
$$
\text{bias}_{pop}(u, i_t) = \text{LogPop}(i_t) - \frac{1}{n} \sum_{k=1}^n \text{LogPop}(i_k),
$$
where the popularity $\text{Pop}(i)$ of an item \(i\) is first computed as the sum of its relevant interactions across all training sequences \( \mathcal{D} \) as: 
$$\text{Pop}(i) = \sum_{s \in \mathcal{D}} C(s, i),
$$
with \(C(s, i)\) representing the number of times item \(i\) appears in sequence \(s\). The logarithm of the popularity is then given by: 
$$ \text{LogPop}(i) = \log(1 + \text{Pop}(i)).$$ Finally, we evaluate the popularity bias by computing the expectation $\mathbb{E}[\text{bias}_{pop}(u, i_t)]$.

\section{Time Efficiency Analysis}\label{ref:time}

We conduct a comparative analysis of training time between S-DPO and NAPO across three datasets: LastFM, Goodreads, and Steam.

\begin{table}[h]
\centering
\caption{Training time comparison between S-DPO and NAPO of the same sampled negatives.}
\label{tab:time_analysis}
\begin{tabular}{cccc}
\toprule
{Method} & {LastFM} & {Goodreads} & {Steam} \\
\midrule
S-DPO & 5.5h & 19.5h & 25h \\
NAPO  & {4.2h} & {15.3h} & {20h} \\
\bottomrule
\end{tabular}
\end{table}

 As shown in Table~\ref{tab:time_analysis}, NAPO consistently outperforms S-DPO in terms of training time, demonstrating a reduction of approximately 15-20\% across all datasets. This improvement is attributed to NAPO's efficient in-batch negative sharing strategy, which avoids the redundant computation of logits for negative samples. Additionally, adopting the SimPO architecture and directly using log probabilities as implicit rewards further enhances efficiency.
 
\section{Related Work}
\subsection{LLMs for Recommendation}
Recent work has explored various paradigms for integrating LLMs into recommendation tasks \cite{DBLP:conf/www/HouHMZ23,sheng2024language,DBLP:conf/sigir/0003CSWC24,DBLP:journals/corr/abs-2408-10159}. Approaches like P5 \cite{DBLP:conf/recsys/Geng0FGZ22} and M6-Rec \cite{DBLP:journals/corr/abs-2205-08084} reformulate recommendation as language modeling, enabling multitask learning and zero-shot generalization. Similarly, LLaRA \cite{DBLP:journals/corr/abs-2312-02445} and TALLRec \cite{DBLP:conf/recsys/BaoZZWF023} align sequential recommendations with behavioral patterns through instruction-tuning. Generative models like PALR \cite{DBLP:journals/corr/abs-2305-07622} and Chat-Rec \cite{gao2023chat} emphasize conversational, personalized recommendations to enhance interaction.

A rising trend incorporates external knowledge and multi-modal data. KAR \cite{DBLP:conf/recsys/XiLLCZZCT0024} enriches user-item interactions with open-world knowledge, while AlphaRec \cite{sheng2024language} applies language embeddings for collaborative filtering. VQ-Rec \cite{DBLP:conf/www/HouHMZ23} addresses domain gaps with vector-quantized representations for cross-domain recommendations. Conversely, DRDT \cite{DBLP:journals/corr/abs-2312-11336} and AgentCF \cite{DBLP:conf/www/ZhangHXSMZLW24} simulate user behaviors using LLMs to generate synthetic training data. Studies of ChatGPT reveal promise in explainable and zero-shot recommendations, though efficiency and accuracy challenges remain \cite{DBLP:journals/corr/abs-2304-10149,DBLP:conf/recsys/DaiSZYSXS0X23}.

DPO has proven effective in aligning LLMs with recommendation tasks by optimizing user preference pairs. Extensions like S-DPO \cite{DBLP:journals/corr/abs-2406-09215} and DMPO \cite{DBLP:journals/corr/abs-2405-16127} enhance preference learning through multiple negative samples. RosePO \cite{DBLP:journals/corr/abs-2410-12519} refines this by mitigating noise and biases. However, the full potential of negative sample space remains underexplored for further performance gains. However, these approaches have yet to fully exploit the negative sample space, a dimension that holds significant potential for further refining and enhancing recommendation performance.

\subsection{Preference Alignment in LLMs}
With the rapid advancement and adoption of LLMs \cite{DBLP:conf/iclr/WeiBZGYLDDL22,DBLP:journals/corr/abs-2110-08207}, aligning them with human preferences has become essential for improving their safety, usability, and control. RLHF is widely used, combining reward modeling with techniques like Proximal Policy Optimization (PPO) to align models with human intent \cite{DBLP:conf/nips/Ouyang0JAWMZASR22,DBLP:conf/nips/ChristianoLBMLA17}. However, due to RLHF's high computational cost, efficient alternatives have emerged. DPO \cite{DBLP:conf/nips/RafailovSMMEF23} simplifies the process by using a cross-entropy objective, eliminating the need for explicit reward modeling. Enhanced versions like $\beta$-DPO \cite{DBLP:journals/corr/abs-2407-08639} and rDPO \cite{DBLP:conf/icml/ChowdhuryKN24} improve parameter tuning and robustness against noisy feedback.

New strategies like SimPO \cite{DBLP:journals/corr/abs-2405-14734} boost efficiency with reference-free rewards, while Iterative Reasoning Preference Optimization \cite{DBLP:journals/corr/abs-2404-19733} refines alignment through iterative training. CPO \cite{DBLP:conf/icml/XuSCTSDM024} advances preference optimization in tasks like machine translation by focusing on curated data. RSO \cite{DBLP:conf/iclr/0002ZJKSLL24} enhances preference pair selection for better policy estimation. Unified frameworks like $\Psi$PO \cite{DBLP:conf/aistats/AzarGPMRVC24} provide both theoretical and practical flexibility across a range of tasks.

\section{Boarder Impact}
While this paper focuses primarily on the technical optimization of LLM-based recommenders, it is essential to consider its broader societal implications. On the positive side, the enhanced preference optimization can lead to more accurate and personalized recommendations, improving user satisfaction and reducing exposure to irrelevant content. This can benefit users by providing more relevant information and reducing the cognitive load associated with information overload.

However, the negative societal impacts should also be acknowledged. The use of LLMs for context-dependent negative sampling inherently involves processing large volumes of user interaction data, raising privacy concerns, especially if sensitive user data is involved. Moreover, the dynamic impact assessment mechanism may inadvertently reinforce existing biases if the model disproportionately penalizes certain types of content or user behaviors.

%% file: neurips_2025.bbl
\begin{thebibliography}{10}

\bibitem{DBLP:journals/corr/abs-2407-21783}
Abhimanyu Dubey, Abhinav Jauhri, Abhinav Pandey, Abhishek Kadian, Ahmad Al{-}Dahle, Aiesha Letman, Akhil Mathur, Alan Schelten, Amy Yang, Angela Fan, Anirudh Goyal, Anthony Hartshorn, Aobo Yang, Archi Mitra, Archie Sravankumar, Artem Korenev, Arthur Hinsvark, Arun Rao, Aston Zhang, Aur{\'{e}}lien Rodriguez, Austen Gregerson, Ava Spataru, Baptiste Rozi{\`{e}}re, Bethany Biron, Binh Tang, Bobbie Chern, Charlotte Caucheteux, Chaya Nayak, Chloe Bi, Chris Marra, Chris McConnell, Christian Keller, Christophe Touret, Chunyang Wu, Corinne Wong, Cristian~Canton Ferrer, Cyrus Nikolaidis, Damien Allonsius, Daniel Song, Danielle Pintz, Danny Livshits, David Esiobu, Dhruv Choudhary, Dhruv Mahajan, Diego Garcia{-}Olano, Diego Perino, Dieuwke Hupkes, Egor Lakomkin, Ehab AlBadawy, Elina Lobanova, Emily Dinan, Eric~Michael Smith, Filip Radenovic, Frank Zhang, Gabriel Synnaeve, Gabrielle Lee, Georgia~Lewis Anderson, Graeme Nail, Gr{\'{e}}goire Mialon, Guan Pang, Guillem Cucurell, Hailey Nguyen, Hannah Korevaar, Hu~Xu, Hugo
  Touvron, Iliyan Zarov, Imanol~Arrieta Ibarra, Isabel~M. Kloumann, Ishan Misra, Ivan Evtimov, Jade Copet, Jaewon Lee, Jan Geffert, Jana Vranes, Jason Park, Jay Mahadeokar, Jeet Shah, Jelmer van~der Linde, Jennifer Billock, Jenny Hong, Jenya Lee, Jeremy Fu, Jianfeng Chi, Jianyu Huang, Jiawen Liu, Jie Wang, Jiecao Yu, Joanna Bitton, Joe Spisak, Jongsoo Park, Joseph Rocca, Joshua Johnstun, Joshua Saxe, Junteng Jia, Kalyan~Vasuden Alwala, Kartikeya Upasani, Kate Plawiak, Ke~Li, Kenneth Heafield, Kevin Stone, and et~al.
\newblock The llama 3 herd of models.
\newblock {\em CoRR}, abs/2407.21783, 2024.

\bibitem{DBLP:journals/corr/abs-2303-08774}
OpenAI.
\newblock {GPT-4} technical report.
\newblock {\em CoRR}, abs/2303.08774, 2023.

\bibitem{DBLP:journals/corr/abs-2305-07622}
Zheng Chen.
\newblock {PALR:} personalization aware llms for recommendation.
\newblock {\em CoRR}, abs/2305.07622, 2023.

\bibitem{DBLP:conf/recsys/XiLLCZZCT0024}
Yunjia Xi, Weiwen Liu, Jianghao Lin, Xiaoling Cai, Hong Zhu, Jieming Zhu, Bo~Chen, Ruiming Tang, Weinan Zhang, and Yong Yu.
\newblock Towards open-world recommendation with knowledge augmentation from large language models.
\newblock In {\em RecSys}, pages 12--22. {ACM}, 2024.

\bibitem{sheng2024language}
Leheng Sheng, An~Zhang, Yi~Zhang, Yuxin Chen, Xiang Wang, and Tat-Seng Chua.
\newblock Language models encode collaborative signals in recommendation.
\newblock 2024.

\bibitem{DBLP:conf/cvpr/LiuLLL24}
Haotian Liu, Chunyuan Li, Yuheng Li, and Yong~Jae Lee.
\newblock Improved baselines with visual instruction tuning.
\newblock In {\em {CVPR}}, pages 26286--26296. {IEEE}, 2024.

\bibitem{DBLP:conf/iclr/WeiBZGYLDDL22}
Jason Wei, Maarten Bosma, Vincent~Y. Zhao, Kelvin Guu, Adams~Wei Yu, Brian Lester, Nan Du, Andrew~M. Dai, and Quoc~V. Le.
\newblock Finetuned language models are zero-shot learners.
\newblock In {\em {ICLR}}. OpenReview.net, 2022.

\bibitem{wang2023generative}
Wenjie Wang, Xinyu Lin, Fuli Feng, Xiangnan He, and Tat-Seng Chua.
\newblock Generative recommendation: Towards next-generation recommender paradigm.
\newblock {\em arXiv preprint arXiv:2304.03516}, 2023.

\bibitem{hou2024large}
Yupeng Hou, Junjie Zhang, Zihan Lin, Hongyu Lu, Ruobing Xie, Julian McAuley, and Wayne~Xin Zhao.
\newblock Large language models are zero-shot rankers for recommender systems.
\newblock In {\em European Conference on Information Retrieval}, pages 364--381. Springer, 2024.

\bibitem{DBLP:conf/nips/RafailovSMMEF23}
Rafael Rafailov, Archit Sharma, Eric Mitchell, Christopher~D. Manning, Stefano Ermon, and Chelsea Finn.
\newblock Direct preference optimization: Your language model is secretly a reward model.
\newblock In {\em NeurIPS}, 2023.

\bibitem{DBLP:conf/uai/RendleFGS09}
Steffen Rendle, Christoph Freudenthaler, Zeno Gantner, and Lars Schmidt{-}Thieme.
\newblock {BPR:} bayesian personalized ranking from implicit feedback.
\newblock In {\em {UAI}}, pages 452--461. {AUAI} Press, 2009.

\bibitem{DBLP:journals/corr/abs-2405-14734}
Yu~Meng, Mengzhou Xia, and Danqi Chen.
\newblock Simpo: Simple preference optimization with a reference-free reward.
\newblock {\em CoRR}, abs/2405.14734, 2024.

\bibitem{DBLP:journals/corr/abs-2406-09215}
Yuxin Chen, Junfei Tan, An~Zhang, Zhengyi Yang, Leheng Sheng, Enzhi Zhang, Xiang Wang, and Tat{-}Seng Chua.
\newblock On softmax direct preference optimization for recommendation.
\newblock {\em CoRR}, abs/2406.09215, 2024.

\bibitem{DBLP:journals/corr/abs-2410-12519}
Jiayi Liao, Xiangnan He, Ruobing Xie, Jiancan Wu, Yancheng Yuan, Xingwu Sun, Zhanhui Kang, and Xiang Wang.
\newblock Rosepo: Aligning llm-based recommenders with human values.
\newblock {\em CoRR}, abs/2410.12519, 2024.

\bibitem{gao2024sprec}
Chongming Gao, Ruijun Chen, Shuai Yuan, Kexin Huang, Yuanqing Yu, and Xiangnan He.
\newblock Sprec: Leveraging self-play to debias preference alignment for large language model-based recommendations.
\newblock {\em arXiv preprint arXiv:2412.09243}, 2024.

\bibitem{DBLP:journals/corr/abs-2405-16127}
Zhuoxi Bai, Ning Wu, Fengyu Cai, Xinyi Zhu, and Yun Xiong.
\newblock Finetuning large language model for personalized ranking.
\newblock {\em CoRR}, abs/2405.16127, 2024.

\bibitem{wu2020sgl}
Jiancan Wu, Xiang Wang, Fuli Feng, Xiangnan He, Liang Chen, Jianxun Lian, and Xing Xie.
\newblock Self-supervised graph learning for recommendation.
\newblock {\em CoRR}, abs/2010.10783, 2020.

\bibitem{wu2024ssm}
Jiancan Wu, Xiang Wang, Xingyu Gao, Jiawei Chen, Hongcheng Fu, and Tianyu Qiu.
\newblock On the effectiveness of sampled softmax loss for item recommendation.
\newblock {\em {ACM} Trans. Inf. Syst.}, 42(4):98:1--98:26, 2024.

\bibitem{DBLP:conf/cikm/MaoZWDDXH21}
Kelong Mao, Jieming Zhu, Jinpeng Wang, Quanyu Dai, Zhenhua Dong, Xi~Xiao, and Xiuqiang He.
\newblock Simplex: {A} simple and strong baseline for collaborative filtering.
\newblock In {\em {CIKM}}, pages 1243--1252. {ACM}, 2021.

\bibitem{ma2024negative}
Haokai Ma, Ruobing Xie, Lei Meng, Fuli Feng, Xiaoyu Du, Xingwu Sun, Zhanhui Kang, and Xiangxu Meng.
\newblock Negative sampling in recommendation: A survey and future directions.
\newblock {\em arXiv preprint arXiv:2409.07237}, 2024.

\bibitem{yang2020mixed}
Ji~Yang, Xinyang Yi, Derek Zhiyuan~Cheng, Lichan Hong, Yang Li, Simon Xiaoming~Wang, Taibai Xu, and Ed~H Chi.
\newblock Mixed negative sampling for learning two-tower neural networks in recommendations.
\newblock In {\em Companion proceedings of the web conference 2020}, pages 441--447, 2020.

\bibitem{zhou2021contrastive}
Chang Zhou, Jianxin Ma, Jianwei Zhang, Jingren Zhou, and Hongxia Yang.
\newblock Contrastive learning for debiased candidate generation in large-scale recommender systems.
\newblock In {\em Proceedings of the 27th ACM SIGKDD Conference on Knowledge Discovery \& Data Mining}, pages 3985--3995, 2021.

\bibitem{bruch2019revisiting}
Sebastian Bruch, Masrour Zoghi, Michael Bendersky, and Marc Najork.
\newblock Revisiting approximate metric optimization in the age of deep neural networks.
\newblock In {\em Proceedings of the 42nd international ACM SIGIR conference on research and development in information retrieval}, pages 1241--1244, 2019.

\bibitem{DBLP:conf/icdm/KangM18}
Wang{-}Cheng Kang and Julian~J. McAuley.
\newblock Self-attentive sequential recommendation.
\newblock In {\em {ICDM}}, pages 197--206. {IEEE} Computer Society, 2018.

\bibitem{DBLP:conf/recsys/2011hetrec}
Iv{\'{a}}n Cantador, Peter Brusilovsky, and Tsvi Kuflik, editors.
\newblock {\em Proceedings of the 2nd International Workshop on Information Heterogeneity and Fusion in Recommender Systems, HetRec '11, Chicago, Illinois, USA, October 27, 2011}. {ACM}, 2011.

\bibitem{DBLP:conf/nips/Ouyang0JAWMZASR22}
Long Ouyang, Jeffrey Wu, Xu~Jiang, Diogo Almeida, Carroll~L. Wainwright, Pamela Mishkin, Chong Zhang, Sandhini Agarwal, Katarina Slama, Alex Ray, John Schulman, Jacob Hilton, Fraser Kelton, Luke Miller, Maddie Simens, Amanda Askell, Peter Welinder, Paul~F. Christiano, Jan Leike, and Ryan Lowe.
\newblock Training language models to follow instructions with human feedback.
\newblock In {\em NeurIPS}, 2022.

\bibitem{DBLP:conf/nips/ChristianoLBMLA17}
Paul~F. Christiano, Jan Leike, Tom~B. Brown, Miljan Martic, Shane Legg, and Dario Amodei.
\newblock Deep reinforcement learning from human preferences.
\newblock In {\em {NIPS}}, pages 4299--4307, 2017.

\bibitem{bradley1952rank}
Ralph~Allan Bradley and Milton~E Terry.
\newblock Rank analysis of incomplete block designs: I. the method of paired comparisons.
\newblock {\em Biometrika}, 39(3/4):324--345, 1952.

\bibitem{DBLP:conf/recsys/BaoZZWF023}
Keqin Bao, Jizhi Zhang, Yang Zhang, Wenjie Wang, Fuli Feng, and Xiangnan He.
\newblock Tallrec: An effective and efficient tuning framework to align large language model with recommendation.
\newblock In {\em RecSys}, pages 1007--1014. {ACM}, 2023.

\bibitem{DBLP:journals/corr/abs-2408-10159}
Xiaoyu Kong, Jiancan Wu, An~Zhang, Leheng Sheng, Hui Lin, Xiang Wang, and Xiangnan He.
\newblock Customizing language models with instance-wise lora for sequential recommendation.
\newblock {\em CoRR}, abs/2408.10159, 2024.

\bibitem{DBLP:journals/corr/abs-2312-02445}
Jiayi Liao, Sihang Li, Zhengyi Yang, Jiancan Wu, Yancheng Yuan, and Xiang Wang.
\newblock Llara: Aligning large language models with sequential recommenders.
\newblock {\em CoRR}, abs/2312.02445, 2023.

\bibitem{DBLP:journals/corr/abs-2106-09685}
Edward~J. Hu, Yelong Shen, Phillip Wallis, Zeyuan Allen{-}Zhu, Yuanzhi Li, Shean Wang, and Weizhu Chen.
\newblock Lora: Low-rank adaptation of large language models.
\newblock {\em CoRR}, abs/2106.09685, 2021.

\bibitem{plackett1975analysis}
Robin~L Plackett.
\newblock The analysis of permutations.
\newblock {\em Journal of the Royal Statistical Society Series C: Applied Statistics}, 24(2):193--202, 1975.

\bibitem{debreu1960individual}
Gerard Debreu.
\newblock Individual choice behavior: A theoretical analysis, 1960.

\bibitem{DBLP:journals/corr/abs-2410-10148}
Junkang Wu, Xue Wang, Zhengyi Yang, Jiancan Wu, Jinyang Gao, Bolin Ding, Xiang Wang, and Xiangnan He.
\newblock {\(\alpha\)}-dpo: Adaptive reward margin is what direct preference optimization needs.
\newblock {\em CoRR}, abs/2410.10148, 2024.

\bibitem{DBLP:conf/nips/VaswaniSPUJGKP17}
Ashish Vaswani, Noam Shazeer, Niki Parmar, Jakob Uszkoreit, Llion Jones, Aidan~N. Gomez, Lukasz Kaiser, and Illia Polosukhin.
\newblock Attention is all you need.
\newblock In {\em {NIPS}}, pages 5998--6008, 2017.

\bibitem{DBLP:journals/corr/abs-2407-08639}
Junkang Wu, Yuexiang Xie, Zhengyi Yang, Jiancan Wu, Jinyang Gao, Bolin Ding, Xiang Wang, and Xiangnan He.
\newblock {\(\beta\)}-dpo: Direct preference optimization with dynamic {\(\beta\)}.
\newblock {\em CoRR}, abs/2407.08639, 2024.

\bibitem{DBLP:journals/ir/ChapelleK10}
Olivier Chapelle and S.~Sathiya Keerthi.
\newblock Efficient algorithms for ranking with svms.
\newblock {\em Inf. Retr.}, 13(3):201--215, 2010.

\bibitem{DBLP:books/daglib/0097035}
Vladimir Vapnik.
\newblock {\em Statistical learning theory}.
\newblock Wiley, 1998.

\bibitem{platt1999probabilistic}
John Platt et~al.
\newblock Probabilistic outputs for support vector machines and comparisons to regularized likelihood methods.
\newblock {\em Advances in large margin classifiers}, 10(3):61--74, 1999.

\bibitem{DBLP:journals/corr/HidasiKBT15}
Bal{\'{a}}zs Hidasi, Alexandros Karatzoglou, Linas Baltrunas, and Domonkos Tikk.
\newblock Session-based recommendations with recurrent neural networks.
\newblock In {\em {ICLR} (Poster)}, 2016.

\bibitem{DBLP:conf/wsdm/TangW18}
Jiaxi Tang and Ke~Wang.
\newblock Personalized top-n sequential recommendation via convolutional sequence embedding.
\newblock In {\em {WSDM}}, pages 565--573. {ACM}, 2018.

\bibitem{DBLP:journals/corr/abs-2307-09288}
Hugo Touvron, Louis Martin, Kevin Stone, Peter Albert, Amjad Almahairi, Yasmine Babaei, Nikolay Bashlykov, Soumya Batra, Prajjwal Bhargava, Shruti Bhosale, Dan Bikel, Lukas Blecher, Cristian Canton{-}Ferrer, Moya Chen, Guillem Cucurull, David Esiobu, Jude Fernandes, Jeremy Fu, Wenyin Fu, Brian Fuller, Cynthia Gao, Vedanuj Goswami, Naman Goyal, Anthony Hartshorn, Saghar Hosseini, Rui Hou, Hakan Inan, Marcin Kardas, Viktor Kerkez, Madian Khabsa, Isabel Kloumann, Artem Korenev, Punit~Singh Koura, Marie{-}Anne Lachaux, Thibaut Lavril, Jenya Lee, Diana Liskovich, Yinghai Lu, Yuning Mao, Xavier Martinet, Todor Mihaylov, Pushkar Mishra, Igor Molybog, Yixin Nie, Andrew Poulton, Jeremy Reizenstein, Rashi Rungta, Kalyan Saladi, Alan Schelten, Ruan Silva, Eric~Michael Smith, Ranjan Subramanian, Xiaoqing~Ellen Tan, Binh Tang, Ross Taylor, Adina Williams, Jian~Xiang Kuan, Puxin Xu, Zheng Yan, Iliyan Zarov, Yuchen Zhang, Angela Fan, Melanie Kambadur, Sharan Narang, Aur{\'{e}}lien Rodriguez, Robert Stojnic, Sergey Edunov,
  and Thomas Scialom.
\newblock Llama 2: Open foundation and fine-tuned chat models.
\newblock {\em CoRR}, abs/2307.09288, 2023.

\bibitem{gao2023chat}
Yunfan Gao, Tao Sheng, Youlin Xiang, Yun Xiong, Haofen Wang, and Jiawei Zhang.
\newblock Chat-rec: Towards interactive and explainable llms-augmented recommender system.
\newblock {\em arXiv preprint arXiv:2303.14524}, 2023.

\bibitem{DBLP:conf/sigir/YuanYSLFYPN23}
Zheng Yuan, Fajie Yuan, Yu~Song, Youhua Li, Junchen Fu, Fei Yang, Yunzhu Pan, and Yongxin Ni.
\newblock Where to go next for recommender systems? {ID-} vs. modality-based recommender models revisited.
\newblock In {\em {SIGIR}}, pages 2639--2649. {ACM}, 2023.

\bibitem{DBLP:conf/www/HouHMZ23}
Yupeng Hou, Zhankui He, Julian~J. McAuley, and Wayne~Xin Zhao.
\newblock Learning vector-quantized item representation for transferable sequential recommenders.
\newblock In {\em {WWW}}, pages 1162--1171. {ACM}, 2023.

\bibitem{DBLP:conf/sigir/0003CSWC24}
An~Zhang, Yuxin Chen, Leheng Sheng, Xiang Wang, and Tat{-}Seng Chua.
\newblock On generative agents in recommendation.
\newblock In {\em {SIGIR}}, pages 1807--1817. {ACM}, 2024.

\bibitem{DBLP:conf/recsys/Geng0FGZ22}
Shijie Geng, Shuchang Liu, Zuohui Fu, Yingqiang Ge, and Yongfeng Zhang.
\newblock Recommendation as language processing {(RLP):} {A} unified pretrain, personalized prompt {\&} predict paradigm {(P5)}.
\newblock In {\em RecSys}, pages 299--315. {ACM}, 2022.

\bibitem{DBLP:journals/corr/abs-2205-08084}
Zeyu Cui, Jianxin Ma, Chang Zhou, Jingren Zhou, and Hongxia Yang.
\newblock M6-rec: Generative pretrained language models are open-ended recommender systems.
\newblock {\em CoRR}, abs/2205.08084, 2022.

\bibitem{DBLP:journals/corr/abs-2312-11336}
Yu~Wang, Zhiwei Liu, Jianguo Zhang, Weiran Yao, Shelby Heinecke, and Philip~S. Yu.
\newblock {DRDT:} dynamic reflection with divergent thinking for llm-based sequential recommendation.
\newblock {\em CoRR}, abs/2312.11336, 2023.

\bibitem{DBLP:conf/www/ZhangHXSMZLW24}
Junjie Zhang, Yupeng Hou, Ruobing Xie, Wenqi Sun, Julian~J. McAuley, Wayne~Xin Zhao, Leyu Lin, and Ji{-}Rong Wen.
\newblock Agentcf: Collaborative learning with autonomous language agents for recommender systems.
\newblock In {\em {WWW}}, pages 3679--3689. {ACM}, 2024.

\bibitem{DBLP:journals/corr/abs-2304-10149}
Junling Liu, Chao Liu, Renjie Lv, Kang Zhou, and Yan Zhang.
\newblock Is chatgpt a good recommender? {A} preliminary study.
\newblock {\em CoRR}, abs/2304.10149, 2023.

\bibitem{DBLP:conf/recsys/DaiSZYSXS0X23}
Sunhao Dai, Ninglu Shao, Haiyuan Zhao, Weijie Yu, Zihua Si, Chen Xu, Zhongxiang Sun, Xiao Zhang, and Jun Xu.
\newblock Uncovering chatgpt's capabilities in recommender systems.
\newblock In {\em RecSys}, pages 1126--1132. {ACM}, 2023.

\bibitem{DBLP:journals/corr/abs-2110-08207}
Victor Sanh, Albert Webson, Colin Raffel, Stephen~H. Bach, Lintang Sutawika, Zaid Alyafeai, Antoine Chaffin, Arnaud Stiegler, Teven~Le Scao, Arun Raja, Manan Dey, M.~Saiful Bari, Canwen Xu, Urmish Thakker, Shanya Sharma, Eliza Szczechla, Taewoon Kim, Gunjan Chhablani, Nihal~V. Nayak, Debajyoti Datta, Jonathan Chang, Mike~Tian{-}Jian Jiang, Han Wang, Matteo Manica, Sheng Shen, Zheng~Xin Yong, Harshit Pandey, Rachel Bawden, Thomas Wang, Trishala Neeraj, Jos Rozen, Abheesht Sharma, Andrea Santilli, Thibault F{\'{e}}vry, Jason~Alan Fries, Ryan Teehan, Stella Biderman, Leo Gao, Tali Bers, Thomas Wolf, and Alexander~M. Rush.
\newblock Multitask prompted training enables zero-shot task generalization.
\newblock {\em CoRR}, abs/2110.08207, 2021.

\bibitem{DBLP:conf/icml/ChowdhuryKN24}
Sayak~Ray Chowdhury, Anush Kini, and Nagarajan Natarajan.
\newblock Provably robust {DPO:} aligning language models with noisy feedback.
\newblock In {\em {ICML}}. OpenReview.net, 2024.

\bibitem{DBLP:journals/corr/abs-2404-19733}
Richard~Yuanzhe Pang, Weizhe Yuan, Kyunghyun Cho, He~He, Sainbayar Sukhbaatar, and Jason Weston.
\newblock Iterative reasoning preference optimization.
\newblock {\em CoRR}, abs/2404.19733, 2024.

\bibitem{DBLP:conf/icml/XuSCTSDM024}
Haoran Xu, Amr Sharaf, Yunmo Chen, Weiting Tan, Lingfeng Shen, Benjamin~Van Durme, Kenton Murray, and Young~Jin Kim.
\newblock Contrastive preference optimization: Pushing the boundaries of {LLM} performance in machine translation.
\newblock In {\em {ICML}}. OpenReview.net, 2024.

\bibitem{DBLP:conf/iclr/0002ZJKSLL24}
Tianqi Liu, Yao Zhao, Rishabh Joshi, Misha Khalman, Mohammad Saleh, Peter~J. Liu, and Jialu Liu.
\newblock Statistical rejection sampling improves preference optimization.
\newblock In {\em {ICLR}}. OpenReview.net, 2024.

\bibitem{DBLP:conf/aistats/AzarGPMRVC24}
Mohammad~Gheshlaghi Azar, Zhaohan~Daniel Guo, Bilal Piot, R{\'{e}}mi Munos, Mark Rowland, Michal Valko, and Daniele Calandriello.
\newblock A general theoretical paradigm to understand learning from human preferences.
\newblock In {\em {AISTATS}}, volume 238 of {\em Proceedings of Machine Learning Research}, pages 4447--4455. {PMLR}, 2024.

\end{thebibliography}
